\documentclass[12pt, a4paper]{article}
\usepackage{graphicx}
\usepackage{subeqnarray}
\begin{document}
%
%                         _________________
%                         |               |
%                         |               |
%                         |  VERSIONE  C  |
%                         |               |
%                         _________________
%
%
%
\def\astrobj#1{#1}
\newenvironment{lefteqnarray}{\arraycolsep=0pt\begin{eqnarray}}
{\end{eqnarray}\protect\aftergroup\ignorespaces}
\newenvironment{lefteqnarray*}{\arraycolsep=0pt\begin{eqnarray*}}
{\end{eqnarray*}\protect\aftergroup\ignorespaces}
\newenvironment{leftsubeqnarray}{\arraycolsep=0pt\begin{subeqnarray}}
{\end{subeqnarray}\protect\aftergroup\ignorespaces}
\newcommand{\diff}{{\rm\,d}}
\newcommand{\img}{{\rm i}}
\newcommand{\appleq}{\stackrel{<}{\sim}}
\newcommand{\appgeq}{\stackrel{>}{\sim}}
\newcommand{\Int}{\mathop{\rm Int}\nolimits}
\newcommand{\Nint}{\mathop{\rm Nint}\nolimits}
\newcommand{\arcsinh}{\mathop{\rm arcsinh}\nolimits}
\newcommand{\range}{{\rm -}}
%newcommand{\erf}{\mathop{\rm erf}\nolimits}
%\newcommand{\psfc}{\mathop{\rm psfc}\nolimits}
%\newcommand{\Psf}{\mathop{\rm psf}\nolimits}
\newcommand{\displayfrac}[2]{\frac{\displaystyle #1}{\displaystyle #2}}
\def\astrobj#1{#1}
%\begin{titlepage}
%\setcounter{page}{0}
%\headnote{Astron.~Nachr.~000 (2001) 0, 000--000}
%\makeheadline
%
\title{The fractional virial potential energy in two-component systems}
\author{{R.~Caimmi}\footnote{
{\it Astronomy Department, Padua Univ., Vicolo Osservatorio 2,
I-35122 Padova, Italy}
email: roberto.caimmi@unipd.it~~~
fax: 39-049-8278212}
, {T.~Valentinuzzi}\footnote{
{\it Astronomy Department, Padua Univ., Vicolo Osservatorio 2,
I-35122 Padova, Italy}
email: tiziano.valentinuzzi@unipd.it~~~
fax: 39-049-8278212}
\phantom{agga}}
%
%\medskip
%\small{Dipartimento di Astronomia}}
%
%\date{Received..................................................
%Accepted..................................................}
\maketitle
\begin{quotation}
\section*{}
\begin{Large}
\begin{center}
%\summary

Abstract

\end{center}
\end{Large}
\begin{small}

\noindent\noindent

Two-component systems are
conceived as macrogases, and the related equation
of state is expressed using the virial theorem for
subsystems, under the restriction of homeoidally
striated density profiles.   Explicit calculations
are performed for a useful reference case and a
few cases of astrophysical interest, both with
and without truncation radius.
Shallower density profiles are found to yield
an equation of state, $\phi=\phi(y,m)$, characterized
(for assigned values of the fractional mass, $m=M_j/
M_i$) by the occurrence of two extremum points, a
minimum and a maximum, as found in an earlier attempt.
Steeper density profiles produce a
similar equation of state, which implies that
a special value of $m$ is related to a critical
curve where the above mentioned extremum points
reduce to a single horizontal inflexion point,
and curves below the critical one show no extremum
points.   The similarity of the isofractional mass
curves to van der Waals' isothermal curves, suggests the
possibility of a phase transition in a bell-shaped
region of the $({\sf O}y\phi)$ plane, where the
fractional truncation radius along a selected direction is
$y=R_j/R_i$, and the fractional virial potential energy is
$\phi=(E_{ji})_{\rm vir}/(E_{ij})_{\rm vir}$.  Further
investigation is devoted to
mass distributions described by Hernquist (1990)
density profiles, for which an additional relation
can be used to represent a sample of $N=16$ elliptical
galaxies (EGs) on the $({\sf O}y\phi)$ plane.
Even if the evolution of elliptical galaxies
and their hosting dark matter (DM) haloes, in the light of the
model, has been characterized by equal fractional
mass, $m$, and equal scaled truncation radius,
or concentration, $\Xi_u=R_u/r_u^\dagger$,
$u=i,j$, still it cannot be considered as strictly
homologous, due to different values of fractional
truncation radii, $y$, or fractional scaling radii,
$y^\dagger=r_j^\dagger/r_i^\dagger$, deduced from
sample objects.

\noindent
{\it keywords -
galaxies: evolution - dark matter: haloes.}
%END
%\end{titlepage}
\end{small}
\end{quotation}

\section{Introduction} \label{intro}

Ordinary fluids (e.g., gases and liquids) may be
bounded by rigid walls which allow particle number
conservation, avoiding evaporation.   Macroscopical
parameters (pressure, density, and temperature)
remain uniform within the box, due to its reduced
dimensions.   On the other hand, astrophysical
fluids (e.g., stars and galaxies) may be
conceived as bounded by ``gravitational'' walls
which violate particle number conservation by
allowing evaporation.   The macroscopical parameters
exhibit gradients because they cannot
remain uniform within the ``gravitational'' box,
due to its large-scale dimensions.

In particular, sufficiently extended celestial
objects show at least two distinct components:
core-envelope for stars, core-halo for elliptical
galaxies, bulge-disk for spiral and lenticular
galaxies, baryonic-nonbaryonic for virialized
(matter) density perturbations, and matter-dark
energy for virialized (matter + dark energy)
density perturbations.   On this basis, an investigation
on two-component astrophysical fluids appears
useful for the comprehension
and the interpretation of what is inferred
from observations.   To this aim, the choice of
the density profiles is of basic importance.
The laws of ideal and real gases were deduced
from ordinary fluids, characterized by uniform
density profiles.   Accordingly, it is
expected that astrophysical fluid laws
are related to the specified density profiles,
and different laws hold for different matter
distributions.

Strictly speaking, a density profile should
be deduced from the distribution function,
or vice versa.   Unfortunately, the determination
of the distribution function is much more
difficult than in one-component
systems, and only a few cases have been
studied in detail at present (e.g., Ciotti,
1996, 1999).
On the other hand, global properties
exhibited by simple density profiles
(with somewhere negative distribution function)
are expected to maintain a similar trend
in dealing with much more complex density
profiles (with nonnegative distribution
function).

As the current attempt is
mainly aimed to explore global properties
instead of local properties, density profiles
shall be selected according to their intrinsic
simplicity, regardless from the physical
meaning of the distribution function.
Configurations described by simple density
profiles could be sufficiently close to their
counterparts described by self-consistent
density profiles, and related results hold
to a first extent.
In any case, the self-consistency of density
profiles with respect to nonnegativity of
the distribution function, can be checked
using a specific theorem (Ciotti and Pellegrini,
1992).

To a first extent, the particle number shall
be supposed to be conserved, which is equivalent
to conceive the boundary of each subsystem as
a perfectly reflecting surface in order to avoid
evaporation.   In this view, a possible choice
of macroscopical parameters is: the fractional
virial potential energy, $\phi$; the fractional truncation
radius, $y$; and the fractional mass, $m$; as done
in a pioneering paper with respect to uniform
density profiles i.e. homogeneous configurations
(Caimmi and Secco, 1990, hereafter quoted as
CS90).   Accordingly, each subsystem is supposed
to be virialized, in the sense that the virial
equations are satisfied by averaging over a
sufficiently long time, and particles move
within a bounded region (e.g., Landau and Lifchitz,
1966, Chap.\,II, \S10; Caimmi, 2007a).
Then virial equilibrium is a necessary (but not
sufficient) condition for dynamical or hydrostatic
equilibrium, which, on the other hand, does not
imply pressureless configurations, as the stress
tensor is related to the kinetic-energy tensor
(e.g., Binney and Tremaine, 1987, Chap.\,4, \S2).

For sake of simplicity, the applications of the
general theory shall be restricted to homeoidally
striated density profiles (e.g., Roberts, 1962;
Caimmi, 1993; Caimmi and Marmo, 2003, hereafter
quoted as CM03).   The larger effects of asphericity
are expected to occurr in homogeneous configurations,
which have widely been investigated (Brosche et al.,
1983; Caimmi et al., 1984; CS90; Caimmi and Secco,
1992).   Focaloidally striated density profiles
involve far larger difficulty (e.g., Caimmi, 1995,
2003).

The current investigation is mainly devoted to
the following points: (i) expression of an equation
of state for two-component 
systems; (ii) description of global properties
deduced from selected density
profiles; and (iii) application to elliptical
galaxies belonging to a restricted sample, to be
represented on the $({\sf O}y\phi)$ plane, for
fiducial values of model parameters.

The work is organized as follows.   The basic
theory of two-component systems
with homeoidally striated density profiles,
is reviewed and extended (to include both cored
and cuspy matter distributions) in Section
\ref{bath}.   The particularization to selected
density profiles, involving explicit expressions,
is made in Section \ref{spec}.   The results and
related global properties are
described and discussed in Section \ref{resu}.
Elliptical galaxies belonging to a restricted
sample, are represented on the $({\sf O}y\phi)$
plane, for fiducial values of model parameters,
in Section \ref{appl},
where some considerations are drawn.   The
concluding remarks are reported in Section \ref{conc}.

\section{Basic theory}\label{bath}

A general theory of two-component matter distributions
has been exhaustively treated in an earlier paper
(Caimmi and Secco, 1992), and the interested reader
is addressed therein and in parent investigations
(MacMillan, 1930, Chap.\,III, \S76; Limber, 1959;
Neutsch, 1979; Brosche et al., 1983; Caimmi et al., 1984)
for further details.   What is relevant for
the current attempt, shall be reviewed and
extended.   If not otherwise stated,
matter distributions should be
conceived as continuous media instead of discrete
particle sets (e.g., Limber, 1959; Caimmi, 2007a).
In the following, general definitions and related
explicit expressions will be provided.   Readers
mainly interested to a simple reference case and
a few cases of astrophysical interest, might
directly go to Section \ref{spec}, while
readers mainly interested to the results and an
application to elliptical galaxies, might directly
go to Sections \ref{resu} and \ref{appl},
respectively.

\subsection{Kinetic-energy and self
potential-energy tensors}\label{kset}

For an assigned density profile, the kinetic-energy
tensor and the kinetic energy read (Binney and
Tremaine, 1987, Chap.\,4, \S4.3):
\begin{leftsubeqnarray}
\slabel{eq:Eka}
&& (E_{\rm kin})_{pq}=\frac12\int_S\rho(x_1,x_2,x_3)v_pv_q\diff^3S~~;
\quad p=1,2,3~~;\quad q=1,2,3~~; \\
\slabel{eq:Ekb}
&& E_{\rm kin}=\frac12\int_S\rho(x_1,x_2,x_3)\sum_{s=1}^3v_s^2\diff^3S~~;
\label{seq:Ek}
\end{leftsubeqnarray}
while the self potential-energy tensor and the
self potential energy read (Chandrasekhar, 1969,
Chap.\,2, \S10):
\begin{leftsubeqnarray}
\slabel{eq:Epa}
&& (E_{\rm sel})_{pq}=\int_S\rho(x_1,x_2,x_3)x_p\frac{\partial{\cal V}}
{\partial x_q}\diff^3S \nonumber \\
&& %\phantom{(E_{\rm sel})_{pq}}
=-\frac12\int_S\rho(x_1,x_2,x_3){\cal V}_
{pq}(x_1,x_2,x_3)\diff^3S~~;\quad p=1,2,3~~;\quad q=1,2,3~~;\qquad \\
\slabel{eq:Epb}
&& E_{\rm sel}=\int_S\rho(x_1,x_2,x_3)\sum_{s=1}^3x_s\frac{\partial{\cal V}}
{\partial x_s}\diff^3S \nonumber \\
&& \phantom{E_{\rm sel}}=-\frac12\int_S\rho(x_1,x_2,x_3){\cal V}
(x_1,x_2,x_3)\diff^3S~~;
\label{seq:Ep}
\end{leftsubeqnarray}
where $\rho$ is the density, $\diff^3S=\diff x_1
\diff x_2\diff x_3$ an infinitesimal volume
element placed at ${\sf P}=(x_1,x_2,x_3)$,
$v_pv_q=\overline{v_pv_q}$ and $v_s^2=\overline
{(v_s^2)}$ are arithmetic means calculated
within $\diff^3S$, ${\cal V}_{pq}$ and ${\cal V}$
are the gravitational tensor potential and
potential, respectively (Chandrasekhar, 1969,
Chap.\,2, \S10):
\begin{leftsubeqnarray}
\slabel{eq:Npota}
&& {\cal V}_{pq}(x_1,x_2,x_3)=G\int_S\rho(x_1^\prime,x_2^\prime,x_3^\prime)
\frac{(x_p^\prime-x_p)(x_q^\prime-x_q)}{\left\vert\overrightarrow{R}-
\overrightarrow{R^\prime}\right\vert^3}\diff^3S^\prime~~; \nonumber \\
&& p=1,2,3~~;\qquad q=1,2,3~~; \\
\slabel{eq:Npotb}
&& {\cal V}(x_1,x_2,x_3)=G\int_S\rho(x_1^\prime,x_2^\prime,x_3^\prime)
\frac{\diff^3S^\prime}{\left\vert\overrightarrow{R}-
\overrightarrow{R^\prime}\right\vert}~~;
\label{seq:Npotp}
\end{leftsubeqnarray}
$G$ being the constant of gravitation, $\overrightarrow
{R}=\overrightarrow{\sf OP}$ and $\overrightarrow
{R^\prime}=\overrightarrow{{\sf OP}^\prime}$,
${\sf P}^\prime=(x_1^\prime,x_2^\prime,x_3^\prime)$,
radius vectors with origin at the centre of inertia.

Conformly to Eq.\,(\ref{seq:Npotp}), the normalization
of the potential satisfies the boundary condition to
be null at infinite distances, and then positive
elsewhere (e.g., MacMillan, 1930, Chap.\,II, \S20;
Chandrasekhar, 1969, Chap.\,3, \S17; Caimmi and Secco,
2003), instead of being null at the centre of inertia,
and then negative elsewhere (e.g., Binney and Tremaine,
1987, Chap.\,2, \S1; Mouri and Taniguchi, 2003).

Let $i$ and $j$ denote the subsystems of a
two-component matter distribution.   The
potential-energy tensor and the potential energy
may be cast into the form (Caimmi and Secco, 1992):
\begin{leftsubeqnarray}
\slabel{eq:Epta}
&& (E_{\rm pot})_{pq}=[(E_i)_{\rm sel}]_{pq}+[(E_{ij})_{\rm int}]_{pq}+
[(E_{ji})_{\rm int}]_{pq}+[(E_j)_{\rm sel}]_{pq}~~;  \\
\slabel{eq:Eptb}
&& E_{\rm pot}=(E_i)_{\rm sel}+(E_{ij})_{\rm int}+
(E_{ji})_{\rm int}+(E_j)_{\rm sel}~~; 
\label{seq:Ept}
\end{leftsubeqnarray}
where the expression of the potential energy
(e.g., MacMillan, 1930, Chap.\,III, \S76) has
been extended to the potential-energy tensor;
in addition, the interaction potential-energy tensor,
$[(E_{uv})_{\rm int}]_{pq}$, and the interaction
potential energy, $(E_{uv})_{\rm int}$, read:
\begin{leftsubeqnarray}
\slabel{eq:Eia}
&& [(E_{uv})_{\rm int}]_{pq}=-\frac12\int_{S_u}\rho_u(x_1,x_2,x_3)
[{\cal V}_v(x_1,x_2,x_3)]_{pq}\diff^3S~~; \nonumber \\
&& u=i,j~~;\qquad v=j,i~~; \\
\slabel{eq:Eib}
&& (E_{uv})_{\rm int}=-\frac12\int_{S_u}\rho_u(x_1,x_2,x_3)
{\cal V}_v(x_1,x_2,x_3)\diff^3S~~; \nonumber \\
&& u=i,j~~;\qquad v=j,i~~;
\label{seq:Ei}
\end{leftsubeqnarray}
which are symmetric with respect to the exchange of
one component with the other:
\begin{leftsubeqnarray}
\slabel{eq:Eisa}
&& [(E_{ij})_{\rm int}]_{pq}=[(E_{ji})_{\rm int})]_{pq}~~; \\
\slabel{eq:Eisb}
&& (E_{ij})_{\rm int}=(E_{ji})_{\rm int}~~;
\label{seq:Eis}
\end{leftsubeqnarray}
for further details refer to earlier attempts
(MacMillan, 1930, Chap.\,III, \S76; Caimmi
and Secco, 1992).

The tidal potential-energy tensor and the
tidal potential energy read (e.g., Caimmi
and Secco, 1992):
\begin{leftsubeqnarray}
\slabel{eq:Eta}
&& [(E_{uv})_{\rm tid}]_{pq}=\int_{S_u}\rho_u(x_1,x_2,x_3)x_p\frac
{\partial{\cal V}_v}{\partial x_q}\diff^3S~~;
\nonumber \\
&& u=i,j~~;\qquad v=j,i~~; \\
\slabel{eq:Etb}
&& (E_{uv})_{\rm tid}=\int_{S_u}\rho_u(x_1,x_2,x_3)\sum_{s=1}^3x_s\frac
{\partial{\cal V}_v}{\partial x_s}\diff^3S~~; \nonumber \\
&& u=i,j~~;\qquad v=j,i~~;
\label{seq:Et}
\end{leftsubeqnarray}
where the tidal potential energy, $(E_{uv})_
{\rm tid}$, may be conceived as the virial of
the $u$-th component in connection with the
tidal field induced by the $v$-th component
(Brosche et al., 1983).

The tensor and the scalar virial theorem for
a single subsystem, within the tidal field
induced by the other one, read (Caimmi et
al., 1984; Caimmi and Secco, 1992):
\begin{leftsubeqnarray}
\slabel{eq:vir1a}
&& 2[(E_u)_{\rm kin}]_{pq}+[(E_u)_{\rm sel}]_{pq}+[(E_{uv})_{\rm tid}]_{pq}
=0~~;\quad u=i,j~~;\quad v=j,i~~;\qquad \\
\slabel{eq:vir1b}
&& 2(E_u)_{\rm kin}+(E_u)_{\rm sel}+(E_{uv})_{\rm tid}=0~~;
\qquad u=i,j~~;\qquad v=j,i~~;
\label{seq:vir1}
\end{leftsubeqnarray}
which is the generalization of previous
results related to one-component systems
(e.g., Chandrasekhar, 1969, Chap.\,II, \S11;
Binney and Tremaine, 1987, Chap.\,4, \S3).
The validity of the relations (Caimmi and
Secco, 1992):
\begin{leftsubeqnarray}
\slabel{eq:Eita}
&& [(E_{ij})_{\rm tid}]_{pq}+[(E_{ji})_{\rm tid}]_{pq}=
[(E_{ij})_{\rm int}]_{pq}+[(E_{ji})_{\rm int})]_{pq}~~; \\
\slabel{eq:Eitb}
&& (E_{ij})_{\rm tid}+(E_{ji})_{\rm tid}=
(E_{ij})_{\rm int}+(E_{ji})_{\rm int}~~;
\label{seq:Eit}
\end{leftsubeqnarray}
implies the following:
\begin{leftsubeqnarray}
\slabel{eq:Eitra}
&& [(E_{uv})_{\rm tid}]_{pq}=[(E_{uv})_{\rm int}]_{pq}+[(E_{uv})_{\rm res}]_
{pq}~~;\quad u=i,j~~;\quad v=j,i~~; \\
\slabel{eq:Eitrb}
&& (E_{uv})_{\rm tid}=(E_{uv})_{\rm int}+(E_{uv})_{\rm res}~~;
\qquad u=i,j~~;\qquad v=j,i~~;
\label{seq:Eitr}
\end{leftsubeqnarray}
where $[(E_{uv})_{\rm res}]_{pq}$ and
$(E_{uv})_{\rm res}$ are the residual
potential-energy tensor and the
residual potential energy, respectively,
which are antisymmetric with regard to
the exchange of one component with the
other:
\begin{leftsubeqnarray}
\slabel{eq:Eraa}
&& [(E_{ij})_{\rm res}]_{pq}=-[(E_{ji})_{\rm res}]_{pq}~~; \\
\slabel{eq:Erab}
&& (E_{ij})_{\rm res}=-(E_{ji})_{\rm res}~~;
\label{seq:Era}
\end{leftsubeqnarray}
for further details refer to earlier attempts
(Caimmi and Secco, 1992; Caimmi, 2007b).

It can be seen from Eqs.\,(\ref{seq:Eis}),
(\ref{seq:Eitr}), and (\ref{seq:Era}),
that the tidal potential-energy tensor,
$[(E_{uv})_{\rm tid}]_{pq}$, and the tidal
potential energy, $(E_{uv})_{\rm tid}$,
are made of a symmetric term, $[(E_{uv})_
{\rm int}]_{pq}$ and $(E_{uv})_{\rm int}$,
and an antisymmetric term, $[(E_{uv})_{\rm
res}]_{pq}$ and $(E_{uv})_{\rm res}$,
respectively, with regard to the exchange
of one component with the other.

The virial theorem in tensor and in scalar
form, expressed by Eqs.\,(\ref{seq:vir1}),
may be cast into the more compact form:
\begin{leftsubeqnarray}
\slabel{eq:vir2a}
&& 2[(E_u)_{\rm kin}]_{pq}+[(E_{uv})_{\rm vir}]_{pq}=0~~;
\qquad u=i,j~~;\qquad v=j,i~~; \\
\slabel{eq:vir2b}
&& 2(E_u)_{\rm kin}+(E_{uv})_{\rm vir}=0~~;
\qquad u=i,j~~;\qquad v=j,i~~;
\label{seq:vir2}
\end{leftsubeqnarray}
where the virial potential-energy tensor,
$[(E_{uv})_{\rm vir}]_{pq}$, and the
virial potential energy, $(E_{uv})_{\rm vir}$,
are defined as:
\begin{leftsubeqnarray}
\slabel{eq:Eva}
&& [(E_{uv})_{\rm vir}]_{pq}=[(E_u)_{\rm sel}]_{pq}+[(E_{uv})_{\rm tid}]_
{pq}~~;\qquad u=i,j~~;\qquad v=j,i~~;\quad \\
\slabel{eq:Evb}
&& (E_{uv})_{\rm vir}=(E_u)_{\rm sel}+(E_{uv})_{\rm tid}~~;
\qquad u=i,j~~;\qquad v=j,i~~;
\label{seq:Ev}
\end{leftsubeqnarray}
where, in general, the virial potential energy is
usually named ``the virial of the system''
(Clausius, 1870).   In the case under
discussion, the ``system'' relates to the
$u$-th component within the tidal potential
induced by the $v$-th component.

For assigned density profiles, the virial
potential-energy tensor and the virial energy
of each subsystem can be determined, together
with their kinetic counterparts via Eqs.\,(\ref
{eq:vir2a}) and (\ref{eq:vir2b}), respectively,
which constrain, in turn, the orbital anisotropy
on each subsystem.   More specifically, the sum
of mean orbital kinetic-energy tensor and kinetic
energy, has to reproduce $[(E_u)_{\rm kin}]_{pq}$
and $(E_u)_{\rm kin}$, respectively.

When a system is not entirely included within
its truncation radius, the usual form of the
virial theorem, $2E_{\rm kin}+E_{\rm pot}=0$,
should be extended as $2E_{\rm kin}+E_{\rm pot}=
3pS$, where the last is a surface term (e.g.,
The and White, 1986; Carlberg et al., 1996;
Girardi et al., 1998).   In dealing with a
specified subsystem, it shall be intended that
no mass exists outside the related truncation
radius, which makes a null surface term.

To avoid the determination of the gravitational
potential, which is the most difficult step
towards an explicit expression of the
potential-energy tensors and potential energies,
a particular procedure shall be followed under
the restrictive assumption of homeoidally
striated density profiles (Roberts, 1962).

\subsection{Homeoidally striated density profiles}
\label{hsdp}

Let the isopycnic (i.e. constant density) surfaces
be defined by the following law (CM03):
\begin{leftsubeqnarray}
\slabel{eq:rhoa}
&& \rho=\rho^\dagger f(\xi)~~;\quad f(1)=1~~; \\
\slabel{eq:rhob}
&& \xi^2=\sum_{\ell=1}^3\frac{x_\ell^2}{(a_\ell^\dagger)^
2}~~;\quad0\le\xi\le\Xi~~;
\label{seq:rho}
\end{leftsubeqnarray}
where $\rho^\dagger=\rho(1)$ , $a_\ell^\dagger$, are
the density and the semiaxes, respectively, of a
reference isopycnic surface, and $\Xi$ corresponds
to the truncation isopycnic surface, related to
semiaxes, $a_\ell$.   The scaled radial coordinate, $\xi$,
and the scaled density, $f$, may be conceived
as the generalization of their counterparts related
to polytropes (e.g., Chandrasekhar, 1939, Chap.\,IV,
\S4; Horedt, 2004, Chap.\,2, \S2.1).

According to Eqs.\,(\ref{seq:rho}), the scaling
density, $\rho^\dagger$, and the
scaling radius,
$r^\dagger$, correspond to a single boundary, which
allows the description of both cored and cuspy density
profiles.   The assumption that the system is
homeoidally striated implies the relation (CM03):
\begin{equation}
\label{eq:csir}
\xi=\frac r{r^\dagger}~~;
\end{equation}
and, in particular:
\begin{equation}
\label{eq:CsiR}
\Xi=\frac R{r^\dagger}~~;
\end{equation}
which, in any case, is independent of the radial coordinate,
$r$, of the generic point on the selected isopycnic
surface.

\subsection{Mass and inertia tensor}\label{mait}

The function (Roberts, 1962):
\begin{leftsubeqnarray}
\slabel{eq:Fa}
&& F(\xi)=2\int_\xi^\Xi f(\xi^\prime)\xi^\prime\diff
\xi^\prime~~; \\
\slabel{eq:Fb}
&& F(\Xi)=0~~;\quad\frac{\diff F}{\diff\xi}=-2\xi
f(\xi)~~; \\
\slabel{eq:Fc}
&& \int_0^\Xi f(\xi)\xi^n\diff\xi=\frac{n-1}2\int_
0^\Xi F(\xi)\xi^{n-2}\diff\xi~~;\qquad n>1~~;
\label{eq:F}
\end{leftsubeqnarray}
allows the calculation of the total mass as
(CM03):
\begin{leftsubeqnarray}
\slabel{eq:Ma}
&& M=\nu_{\rm mas}M^\dagger~~; \\
\slabel{eq:Mb}
&& \nu_{\rm mas}=\frac32\int_0^\Xi F(\xi)\diff\xi~~; \\
\slabel{eq:Mc}
&& M^\dagger=\frac{4\pi}3\rho^\dagger a_1^\dagger a_2^\dagger a_3^\dagger~~;
\label{seq:M}
\end{leftsubeqnarray}
and the inertia tensor as (CM03):
\begin{leftsubeqnarray}
\slabel{eq:Ia}
&& I_{pq}=\delta_{pq}\nu_{\rm inr}M^\dagger(a_p^\dagger)^2~~; \\
\slabel{eq:Ib}
&& \nu_{\rm inr}=\frac32\int_0^\Xi F(\xi)\xi^2\diff\xi~~;
\label{seq:I}
\end{leftsubeqnarray}
where the coefficients, $\nu_{\rm mas}$ and $\nu_{\rm inr}$, are
shape-independent and may be conceived as profile
factors (Caimmi, 1993), and $\delta_{pq}$ is the
Kronecker symbol.   For a different formulation
of the inertia tensor refer to earlier attempts
(e.g., Landau and Lifchitz, 1966, Chap.\,VI,
\S32; Bett et al., 2007).

The mass within an isopycnic surface, $\rho=
\rho^\dagger f(\xi)$, is
$M(\xi)=\nu_{\rm mas}(\xi)M^\dagger$, where
$\nu_{\rm mas}(\xi)$ is expressed by Eq.\,(\ref
{eq:Mb}) with $\xi$ instead of $\Xi$.   The
related mean density is $\overline{\rho}(\xi)=
M(\xi)/S(\xi)$, where $S(\xi)$ is the volume
bounded by the isopycnic surface.

\subsection{Potential-energy tensors and potential energies}
\label{pete}

The self potential-energy tensor and the self
potential energy read (CM03):
\begin{leftsubeqnarray}
\slabel{eq:Esa}
&& (E_{\rm sel})_{pq}=-\delta_{pq}\nu_{\rm sel}
\frac{G(M^\dagger)^2}{a^\dagger_1}B_p~~; \\
\slabel{eq:Esb}
&& E_{\rm sel}=-\nu_{\rm sel}
\frac{G(M^\dagger)^2}{a^\dagger_1}B~~; \\
\slabel{eq:Esc}
&& \nu_{\rm sel}=\frac9{16}\int_0^\Xi F^2(\xi)\diff\xi~~; \\
\slabel{eq:Esd}
&& B_p=\epsilon_{p2}\epsilon_{p3}\int_0^{+\infty}(1+s^\prime)^{-3/2}
(1+\epsilon_{pq}^2s^\prime)^{-1/2}(1+\epsilon_{pr}^2s^\prime)^{-1/2}
\diff s^\prime~~; \\
\slabel{eq:Ese}
&& B=\sum_{s=1}^3B_p~~;\qquad\epsilon_{mn}=\frac{a_m}{a_n}~~;
\label{seq:Es}
\end{leftsubeqnarray}
where $\nu_{\rm sel}$ is a profile factor,
$\epsilon_{mn}$ are axis ratios, and $B_p$
are profile factors which, {\it ipso facto},
depend on the axis ratios only (Caimmi, 1992).

The above results are related to a single
subsystem: strictly speaking, all the
quantities defined in the current section
should be labelled by the index, $u$, in
connection with the $u$-th subsystem (e.g.,
Caimmi and Secco, 1992), but it has been
omitted to gain clarity.   On the other
hand, the formulation of other
potential-energy tensors and potential
energies necessarily involves (at least)
two components.   The related calculations
are very difficult in the general case,
and for this reason only the special
situation of similar and similarly
placed boundaries shall be considered.

With this restriction, the combination
of Eqs.\,(\ref{eq:csir}), (\ref{eq:CsiR}),
and (\ref{eq:Mc}), related to both the
density profiles, the inner to be denoted
as $i$ and the outer as $j$, yields:
\begin{leftsubeqnarray}
\slabel{eq:yma}
&& \xi_i=y^\dagger\xi_j~~;\qquad\frac{\Xi_j}{\Xi_i}=\frac y{y^\dagger}~~;
\qquad\frac{(\nu_j)_{\rm mas}}{(\nu_i)_{\rm mas}}=\frac m{m^\dagger}~~; \\
\slabel{eq:ymb}
&& y=\frac{R_j}{R_i}~~;\qquad y^\dagger=\frac{r_j^\dagger}{r_i^\dagger}~~;
\qquad m=\frac{M_j}{M_i}~~;\qquad m^\dagger=\frac{M_j^\dagger}{M_i^\dagger}~~;
\label{seq:ym}
\end{leftsubeqnarray}
which allows the expression of the other
potential-energy tensors and potential
energies, as (CM03):
\begin{leftsubeqnarray}
\slabel{eq:Exxxa}
&& [(E_{uv})_{\rm xxx}]_{pq}=-\delta_{pq}
\frac{G(M_u^\dagger)^2}{(a_u^\dagger)_1}(\nu_{uv})_{\rm xxx}B_p~~; \\
\slabel{eq:Exxxb}
&& (E_{uv})_{\rm xxx}=-
\frac{G(M_u^\dagger)^2}{(a_u^\dagger)_1}(\nu_{uv})_{\rm xxx}B~~; \\
\slabel{eq:Exxxc}
&& u=i,j~~;\qquad v=j,i~~;\qquad{\rm xxx}={\rm int},\,{\rm tid},\,
{\rm res},\,{\rm vir}~~;
\label{seq:Exxx}
\end{leftsubeqnarray}
and the explicit expression of the
profile factors reads (CM03):
\begin{leftsubeqnarray}
\slabel{eq:nuuva}
&& (\nu_{ij})_{\rm int}=(\nu_{ji})_{\rm int}=-\frac9{16}m^\dagger
\left[w^{({\rm int})}(\eta)+w^{({\rm ext})}(\eta)\right]~~; \\
\slabel{eq:nuuvb}
&& (\nu_{ij})_{\rm tid}=-\frac98m^\dagger w^{({\rm ext})}(\eta)~~;\qquad
(\nu_{ji})_{\rm tid}=-\frac98\frac{y^\dagger}{m^\dagger}w^{({\rm int})}
(\eta)~~; \\
\slabel{eq:nuuvc}
&& (\nu_{ij})_{\rm res}=-(\nu_{ji})_{\rm res}=-\frac9{16}m^\dagger
\left[w^{({\rm int})}(\eta)-w^{({\rm ext})}(\eta)\right]~~; \\
\slabel{eq:nuuvd}
&& (\nu_{uv})_{\rm vir}=(\nu_u)_{\rm sel}+(\nu_{uv})_{\rm tid}~~;\qquad
u=i,j~~;\qquad v=j,i~~; \\
\slabel{eq:nuuve}
&& \eta=\frac{\Xi_i}{y^\dagger}=\frac{\Xi_j}y~~;
\label{seq:nuuv}
\end{leftsubeqnarray}
where the functions, $w^{({\rm int})}$ and
$w^{({\rm ext})}$, are defined as (CM03):
\begin{leftsubeqnarray}
\slabel{eq:wiea}
&& w^{({\rm int})}(\eta)=\int_0^\eta F_j(\xi_j)\frac
{\diff F_i}{\diff\xi_j}\xi_j\diff\xi_j~~; \\
\slabel{eq:wieb}
&& w^{({\rm ext})}(\eta)=\int_0^\eta F_i(\xi_i)\frac
{\diff F_j}{\diff\xi_j}\xi_j\diff\xi_j~~;
\label{seq:wie}
\end{leftsubeqnarray}
in conclusion, Eqs.\,(\ref{seq:Es})-(\ref
{seq:wie}) allow the calculation of the
potential-energy tensors and potential
energies for homeoidally striated density
profiles related to 
similar and similarly placed boundaries.

In the case under discussion, due to
Eqs.\,(\ref{seq:Exxx}), the fractional
virial potential-energy tensor component
equals the fractional virial potential
energy, as:
\begin{equation}
\label{eq:phic}
\phi=\frac{[(E_{ji})_{\rm vir}]_{pq}}{[(E_{ij})_{\rm vir}]_{pq}}=
\frac{(E_{ji})_{\rm vir}}{(E_{ij})_{\rm vir}}=\frac{(m^\dagger)^2}
{y^\dagger}\frac{(\nu_{ji})_{\rm vir}}{(\nu_{ij})_{\rm vir}}~~;
\end{equation}
which, for assigned density profiles,
depends on the reference fractional
mass, $m^\dagger$, and the fractional
scaling radius, $y^\dagger$,
according to Eqs.\,(\ref{seq:ym})-(\ref
{seq:wie}).   The substitution of
Eqs.\,(\ref{seq:ym}) into (\ref{eq:phic})
yields:
\begin{equation}
\label{eq:phi}
\phi=\frac{m^2}y\frac{\Xi_j}{\Xi_i}\left[\frac{(\nu_i)_{\rm mas}}
{(\nu_j)_{\rm mas}}\right]^2\frac{(\nu_{ji})_{\rm vir}}{(\nu_{ij})_
{\rm vir}}~~;
\end{equation}
which, for assigned density profiles,
depends on the fractional mass, $m$,
and the fractional truncation radius, $y$,
according to Eqs.\,(\ref{seq:ym})-(\ref
{seq:wie}).

Strictly speaking, Eqs.\,(\ref{seq:ym})-(\ref
{eq:phi}) are valid provided the indices,
$i$ and $j$, denote the embedded and the
embedding subsystem, respectively, which
implies $y\ge1$.   This is why the above
procedure is valid only for the inner
component, where the Poisson equation
instead of the Laplace equation holds
with respect to the gravitational
potential.   The remaining results,
related to the outer component, are
obtained by use of the symmetry of
the interaction potential-energy tensor
and interaction potential energy,
Eqs.\,(\ref{seq:Eis}), and the
antisymmetry of the residual
potential-energy tensor and residual
potential energy, Eqs.\,(\ref{seq:Era}).

If the role of the two subsystems is
reversed, $0\le y\le1$, it has to be
kept in mind that, in this case, the
inner and the outer component are
denoted by the indices, $j$ and $i$,
respectively, and the reversion must
be done in Eqs.\,(\ref{seq:ym})-(\ref
{eq:phi}), through the following
steps: (i) make the changes: $m\to
m^{-1}$; $m^\dagger\to(m^\dagger)^
{-1}$; $y\to y^{-1}$; $y^\dagger\to
(y^\dagger)^{-1}$;
$\Xi_i\leftrightarrow\Xi_j$;
$(\nu_i)_{\rm mas}\leftrightarrow
(\nu_j)_{\rm mas}$; $(\nu_i)_{\rm
sel}\leftrightarrow(\nu_j)_{\rm sel}$;
(ii) for assigned $y\ge1$ and
$y^\dagger=(\Xi_i/\Xi_j)y$,
calculate the functions, $w^{({\rm
int})}(\eta)$ and $w^{({\rm ext})}
(\eta)$; (iii) calculate the profile
factors, $(\nu_{ij})_{\rm tid}$ and
$(\nu_{ji})_{\rm tid}$, and the
remaining ones when needed; (iv)
calculate the fractional virial
potential energy, $\phi$; (v) make
the changes, $m\to
m^{-1}$; $m^\dagger\to(m^\dagger)^
{-1}$; $y\to y^{-1}$; $y^\dagger\to
(y^\dagger)^{-1}$; $\phi\to\phi^
{-1}$; which allow the extension
of the fractional virial potential
energy, $\phi=(E_{ji})_{\rm vir}/
(E_{ij})_{\rm vir}$, to the domain,
$0\le y\le1$.

In absence of truncation radius,
$\Xi\to+\infty$, $\eta\to+\infty$,
the reversion occurs when the density
drops to zero and nothing changes
except in infinitesimal terms of
higher order and infinite terms of
lower order.   Accordingly, there
is no need to perform the reversion
in this case.

The relation, $\phi=\phi(y^\dagger,
m^\dagger)$, or its counterpart,
$\phi=\phi(y,m)$, expressed by
Eqs.\,(\ref{eq:phic}) and (\ref
{eq:phi}), respectively, may be
conceived as an equation of state
for two-component systems with
assigned homeoidally striated
density profiles.   At the price
of a major complexity, it can be
formulated for any kind of
two-component systems,
hereafter quoted as
``two-component macrogases'',
or more shortly as ``macrogases''.
%In the next section, the
%macrogases equation of state shall
%be determined for a simple but
%unrealistic density profile, to
%be taken as a reference case,
%and a few density profiles of
%astrophysical interest.

\section{Special cases}\label{spec}

The explicit expression of the macrogases
equation of state is, in general, rather
cumbersome and numerical computations
should be preferred to this aim.   On
the other hand, the related procedure is
conceptually simple, as the functions,
$w^{({\rm int})}(\eta)$ and $w^{({\rm ext})}
(\eta)$, and the fractional virial potential
energy, $\phi$, may be calculated using
Eqs.\,(\ref{seq:Es})-(\ref{eq:phi}).

Aiming to provide a description of the
general trend and related features, a
limited number of simple possibilities
shall be analysed in detail, selecting
density profiles from the family:
\begin{equation}
\label{eq:ex3}
f(\xi)=\frac{2^\chi}{\xi^\gamma(1+\xi^\alpha)^\chi}~~;\qquad\chi=
\frac{\beta-\gamma}\alpha~~;
\end{equation}
which is defined by three parameters,
$(\alpha,\beta,\gamma)$.   For
further details refer to earlier
attempts (e.g., Hernquist, 1990;
Zhao, 1996; Caimmi and Marmo, 2004;
Caimmi et al., 2005; Caimmi, 2006b),
where special cases were fully
investigated.

In the following, the macrogases
equation of state shall
be determined for a simple but
unrealistic density profile, to
be taken as a reference case,
and a few density profiles of
astrophysical interest.   The
reader whose attention is
mainly directed to the results
and/or the astrophysical applications
of the model, is free to jump
directly to Section \ref{resu}
and/or \ref{appl}, respectively.

\subsection{UU macrogases}\label{UU}

The related density profiles maintain
uniform, $(\alpha,\beta,\gamma)=(0,0,0)$,
and Eq.\,(\ref{eq:ex3}) reduces to:
\begin{equation}
\label{eq:fU}
f_u(\xi_u)=1~~;\qquad0\le\xi_u\le\Xi_u~~;\qquad u=i,j~~;
\end{equation}
which is equivalent to polytropes with index,
$n=0$ (e.g., Chandrasekhar, 1939, Chap.\,IV,
\S4; Caimmi, 1986) but implies unphysical
situations for stellar fluids (Vandervoort, 1980).
The particularization of Eqs.\,(\ref
{eq:Fa}), (\ref{eq:Mb}), (\ref{eq:Esc}), (\ref
{eq:wiea}), and (\ref{eq:wieb}) to the case of
interest yields:
\begin{lefteqnarray}
\label{eq:FU}
&& F_u(\xi_u)=\Xi_u^2-\xi_u^2~~;\qquad u=i,j~~; \\
\label{eq:numU}
&& (\nu_u)_{\rm mas}=\Xi_u^3~~;\qquad u=i,j~~; \\
\label{eq:nusU}
&& (\nu_u)_{\rm sel}=\frac3{10}\Xi_u^5~~;\qquad u=i,j~~; \\
\label{eq:wiU}
&& w^{({\rm int})}(\eta)=-\frac4{15}\Xi_i^2\eta^3
\left(\frac52y^2-\frac32\right)~~; \\
\label{eq:weU}
&& w^{({\rm ext})}(\eta)=-\frac4{15}\Xi_i^2\eta^3~~;
\end{lefteqnarray}
and, using Eqs.\,(\ref{eq:nuuvb}) and
(\ref{eq:nuuve}), the UU
macrogases equation of state follows from
the particularization of Eq.\,(\ref{eq:phi})
to the case of interest, as:
\begin{equation}
\label{eq:phlU}
\phi=\frac{(m^\dagger)^2}{y^\dagger}\left(\frac y{y^\dagger}\right)^5
\displayfrac{1+\frac{({y^\dagger})^3}{(m^\dagger)}\frac1{y^5}\left(
\frac52y^2-\frac32\right)}{1+\frac{m^\dagger}{(y^\dagger)^3}}~~;\qquad
y\ge1~~;
\end{equation}
and the extension of the above function
to the domain, $0\le y\le1$, following
the procedure outlined in Subsection
\ref{pete}, yields:
\begin{equation}
\label{eq:phuU}
\phi=\left(\frac y{y^\dagger}\right)^5
\displayfrac{m^\dagger(y^\dagger)^2\left[1+\frac{m^\dagger}{(y^\dagger)^3}
\right]}{1+m^\dagger\left(\frac y{y^\dagger}\right)^3\left(\frac52-\frac32
y^2\right)}~~;\qquad0\le y\le1~~;
\end{equation}
where it can be seen that in the special case,
$y=1$, Eqs.\,(\ref{eq:phlU}) and (\ref
{eq:phuU}) do coincide.   If, in addition,
$\Xi_i=\Xi_j$, which implies $y^\dagger=y$,
$m^\dagger=m$, via Eqs.\,(\ref{seq:M}) and
(\ref{seq:ym}), then Eqs.\,(\ref{eq:phlU})
and (\ref{eq:phuU}) reduce to:
\begin{equation}
\label{eq:ph1U}
\phi=m=m^\dagger~~;\qquad y=y^\dagger=1~~;
\end{equation}
or $\phi(1,m)=m$.

Owing to Eqs.\,(\ref{eq:phlU}) and (\ref
{eq:phuU}), the fractional virial potential
energy, $\phi$, is independent of the
fractional truncation radii, $(\Xi_i,\Xi_j)$.

\subsection{PP macrogases}\label{PP}

The related density profiles (Schuster,
1883; Plummer, 1911) imply
$(\alpha,\beta,\gamma)=(2,5,0)$,
and Eq.\,(\ref{eq:ex3}) reduces to:
\begin{equation}
\label{eq:fP}
f_u(\xi_u)=\frac{2^{5/2}}{(1+\xi_u^2)^{5/2}}~~;
\qquad0\le\xi_u\le\Xi_u~~;\qquad u=i,j~~;
\end{equation}
which is equivalent to polytropes with index,
$n=5$; for a formal demonstration, see
Appendix \ref{a:pol5}.
For one-component
systems, hydrostatic equilibrium holds
regardless from the nature of the fluid
(Vandervoort, 1980).

The particularization of Eqs.\,(\ref{eq:Fa}),
(\ref{eq:Mb}), (\ref{eq:Esc}), (\ref{eq:wiea}),
and (\ref{eq:wieb}) to the case of interest
yields:
\begin{lefteqnarray}
\label{eq:FP}
&& F_u(\xi_u)=\frac{2^{7/2}}3\left[\frac1{(1+\xi_u^2)^{3/2}}-\frac1
{(1+\Xi_u^2)^{3/2}}\right]~~;\qquad u=i,j~~; \\
\label{eq:numP}
&& (\nu_u)_{\rm mas}=\frac{2^{5/2}\Xi_u^3}{(1+\Xi_u^2)^{3/2}}~~;\qquad
u=i,j~~; \\
\label{eq:nusP}
&& (\nu_u)_{\rm sel}=\frac{\Xi_u(3\Xi_u^4-8\Xi_u^2+13)}{(1+\Xi_u^2)^3}+
3\arctan\Xi_u~~;\qquad u=i,j~~;
\end{lefteqnarray}
\begin{leftsubeqnarray}
\slabel{eq:wiPa}
&& w^{({\rm int})}(\eta)=-\frac{128}3(y^\dagger)^2\left\{\frac{[(y^\dagger)^
2+7]\img E(y^\dagger,\img\alpha)+4[(y^\dagger)^2-1]\img F(y^\dagger,\img
\alpha)}{3[(y^\dagger)^2-1]^3}\right. \nonumber \\
&& +\left.\frac{P^{(\rm int)}(\eta,y^\dagger)}{3[(y^\dagger)^2-1]^3(\eta^2+1)
[(y^\dagger)^2\eta^2+1]^2}-\frac1{(1+\Xi_j^2)^{3/2}}\frac{\eta^3}{3[(y^
\dagger)^2\eta^2+1]^{3/2}}\right\}~~;\qquad \\
\slabel{eq:wiPb}
&& P^{(\rm int)}(\eta,y^\dagger)=\eta\sqrt{\eta^2+1}\sqrt{(y^\dagger)^2\eta^2
+1}  \nonumber \\
&& \times[(y^\dagger)^6\eta^2(\eta^2+1)+(y^\dagger)^4\eta^2(7\eta^2-4)+
(y^\dagger)^2(11\eta^2+5)+3]~~; \\
\slabel{eq:wiPc}
&& \alpha=\arcsinh\eta~~;\qquad y^\dagger\ne1~~; \\
\slabel{eq:wiPd}
&& w^{({\rm int})}(\eta)=-\frac{128}3(y^\dagger)^2\left[\frac
{\arctan\eta}{16}+\frac{\eta(3\eta^4+8\eta^2-3)}{48(\eta^2+1)^3}\right.
\nonumber \\
&& \left.-\frac1{(1+\Xi_j^2)^{3/2}}
\frac{\eta^3}{3(\eta^2+1)^{3/2}}\right]~~;\qquad y^\dagger=1~~;
\label{seq:wiP}
\end{leftsubeqnarray}
\begin{leftsubeqnarray}
\slabel{eq:wePa}
&& w^{({\rm ext})}(\eta)=-\frac{128}3\left\{-\frac{[7(y^\dagger)^
2+1]\img E(y^\dagger,\img\alpha)+[3(y^\dagger)^4-2(y^\dagger)^2-1]\img
F(y^\dagger,\img\alpha)}{3[(y^\dagger)^2-1]^3}\right. \nonumber \\
&& -\left.\frac{P^{(\rm ext)}(\eta,y^\dagger)}{3[(y^\dagger)^2-1]^3(\eta^2+1)
^2[(y^\dagger)^2\eta^2+1]}-\frac1{(1+\Xi_i^2)^{3/2}}\frac{\eta^3}{3(\eta^2+1)
^{3/2}}\right\}~~;\qquad \\
\slabel{eq:wePb}
&& P^{(\rm ext)}(\eta,y^\dagger)=\eta\sqrt{\eta^2+1}\sqrt{(y^\dagger)^2\eta^2
+1}  \nonumber \\
&& \times[(y^\dagger)^4(7\eta^4+11\eta^2+3)+(y^\dagger)^2(\eta^4+4\eta^2+5)
+\eta^2]~~; \\
\slabel{eq:wePc}
&& \alpha=\arcsinh\eta~~;\qquad y^\dagger\ne1~~; \\
\slabel{eq:wePd}
&& w^{({\rm ext})}(\eta)=-\frac{128}3\left[\frac
{\arctan\eta}{16}+\frac{\eta(3\eta^4+8\eta^2-3)}{48(\eta^2+1)^3}\right.
\nonumber \\
&& \left.-\frac1{(1+\Xi_i^2)^{3/2}}
\frac{\eta^3}{3(\eta^2+1)^{3/2}}\right]~~;\qquad y^\dagger=1~~;
\label{seq:weP}
\end{leftsubeqnarray}
where $\img$ is the imaginary unit, $F$
and $E$ are incomplete elliptic integrals
of the first and second kind, respectively,
defined as:
\begin{leftsubeqnarray}
\slabel{eq:ella}
&& F(k,\beta)=\int_0^\beta\frac{\diff\theta}{\sqrt{1-k^2\sin^2\theta}}=
\int_0^x\frac{\diff t}{\sqrt{1-t^2}\sqrt{1-k^2t^2}}~~; \\
\slabel{eq:ellb}
&& E(k,\beta)=\int_0^\beta\sqrt{1-k^2\sin^2\theta}\diff\theta=
\int_0^x\frac{\sqrt{1-k^2t^2}}{\sqrt{1-t^2}}\diff t~~; \\
\slabel{eq:ellc}
&& x=\sin\beta~~;\qquad t=\sin\theta~~;
\label{seq:ell}
\end{leftsubeqnarray}
for further details refer to specialized
textbooks (e.g., Spiegel, 1968, Chap.\,4,
\S\S34.1-4).

Using Eqs.\,(\ref{eq:nuuvb}) and (\ref
{eq:numP})-(\ref{seq:weP}), the PP
macrogases equation of state is obtained
from the particularization of Eq.\,(\ref
{eq:phi}) to the case of interest for
the domain, $y\ge1$.   The extension to
the domain, $0\le y\le1$, can be done
following the procedure outlined in
Subsection \ref{pete}.

In absence of truncation radius, the density
drops to zero when the radius goes to
infinity, $\Xi\to+\infty$, $\eta\to+\infty$,
and Eqs.\,(\ref{eq:FP})-(\ref{seq:weP})
reduce to:
\begin{lefteqnarray}
\label{eq:FlP}
&& \lim_{\Xi_u\to+\infty}F_u(\xi_u)=\frac{2^{7/2}}3\frac1{(1+\xi_u^2)^{3/2}}~~;
\qquad u=i,j~~; \\
\label{eq:lumP}
&& \lim_{\Xi_u\to+\infty}(\nu_u)_{\rm mas}=2^{5/2}~~;\qquad u=i,j~~; \\
\label{eq:lusP}
&& \lim_{\Xi_u\to+\infty}(\nu_u)_{\rm sel}=\frac{3\pi}2~~;\qquad u=i,j~~;
\end{lefteqnarray}
\begin{leftsubeqnarray}
\slabel{eq:wilPa}
&& \lim_{\eta\to+\infty}w^{({\rm int})}(\eta)=-\frac{128}3(y^\dagger)^2\frac
{[(y^\dagger)^2+7]E(k,\pi/2)-[5(y^\dagger)^2+3]F(k,\pi/2)}{3[(y^\dagger)^2-1]
^3}~~; \nonumber \\
&& k=\sqrt{1-(y^\dagger)^2}~~;\qquad y^\dagger\ne1~~; \\
\slabel{eq:wilPb}
&& \lim_{\eta\to+\infty}w^{({\rm int})}(\eta)=-\frac{4\pi}3~~;\qquad
y^\dagger=1~~;
\label{seq:wilP}
\end{leftsubeqnarray}
\begin{leftsubeqnarray}
\slabel{eq:welPa}
&& \lim_{\eta\to+\infty}w^{({\rm ext})}(\eta)=-\frac{128}3y^\dagger\frac
{[3(y^\dagger)^2+5]F(k,\pi/2)-[7(y^\dagger)^2+1]E(k,\pi/2)}{3[(y^\dagger)^2-1]
^3}~~; \nonumber \\
&& k=\sqrt{1-1/(y^\dagger)^2}~~;\qquad y^\dagger\ne1~~; \\
\slabel{eq:welPb}
&& \lim_{\eta\to+\infty}w^{({\rm ext})}(\eta)=-\frac{4\pi}3~~;\qquad
y^\dagger=1~~;
\label{seq:welP}
\end{leftsubeqnarray}
where, in particular, the related expression
of the tidal energy coincides with its
counterpart calculated in an earlier attempt
(Valentinuzzi, 2006, Chap.\,4, \S4.1).
Using Eqs.\,(\ref{eq:nuuvb}) and
(\ref{eq:lumP})-(\ref{seq:welP}), the PP
macrogases equation of state in the
special situation under discussion, is
obtained from the particularization of
Eq.\,(\ref{eq:phi}) to the case of interest,
for the domain, $y\ge0$.

If, in addition, $y=y^\dagger=1$,
$\Xi_j=\Xi_i$, the combination of
Eqs.\,(\ref{eq:phi}) and (\ref
{eq:lumP})-(\ref{seq:welP}) yields
Eq.\,(\ref{eq:ph1U}).

\subsection{HH macrogases}\label{HH}

The related density profiles (Hernquist,
1990) imply $(\alpha,\beta,\gamma)=(1,4,1)$,
and Eq.\,(\ref{eq:ex3}) reduces to:
\begin{equation}
\label{eq:fH}
f_u(\xi_u)=\frac8{\xi_u(1+\xi_u)^3}~~;
\qquad0\le\xi_u\le\Xi_u~~;\qquad u=i,j~~;
\end{equation}
which  has been proved to be consistent with
nonnegative distribution functions,
in the parameter range of interest
(Ciotti, 1996).

The particularization of Eqs.\,(\ref{eq:Fa}),
(\ref{eq:Mb}), (\ref{eq:Esc}), (\ref{eq:wiea}),
and (\ref{eq:wieb}) to the case of interest
yields:
\begin{lefteqnarray}
\label{eq:FH}
&& F_u(\xi_u)=\frac8{(1+\xi_u)^2}-\frac8{(1+\Xi_u)^2}~~;\qquad u=i,j~~; \\
\label{eq:numH}
&& (\nu_u)_{\rm mas}=\frac{12\Xi_u^2}{(1+\Xi_u)^2}~~;\qquad u=i,j~~; \\
\label{eq:nusH}
&& (\nu_u)_{\rm sel}=\frac{12\Xi_u^3(4+\Xi_u)}{(1+\Xi_u)^4}~~;\qquad
u=i,j~~;
\end{lefteqnarray}
\begin{leftsubeqnarray}
\slabel{eq:wiHa}
&& w^{({\rm int})}(\eta)=-128y^\dagger\left\{\frac12\frac1{(y^\dagger-1)^4}
\left[-\frac{(y^\dagger-1)^2y^\dagger\eta}{(y^\dagger\eta+1)^2}+\frac
{2(y^\dagger-1)\eta}{1+\eta}\right.\right. \nonumber \\
&& \left.+\frac{(y^\dagger-1)(y^\dagger+3)y^\dagger\eta}{y^\dagger\eta+
1}+2(2y^\dagger+1)\ln\frac{\eta+1}{y^\dagger\eta+1}\right] \nonumber \\
&& \left.-\frac12\frac1{(1+\Xi_j)^2}\frac1{(y^\dagger)^2}\left[1-\frac
{2y^\dagger\eta+1}{(y^\dagger\eta+1)^2}\right]\right\}~~;\qquad y^\dagger
\ne1~~; \\
\slabel{eq:wiHb}
&& w^{({\rm int})}(\eta)=-128\left\{\frac1{12}\left[-\frac{4\eta+1}
{(\eta+1)^4}+1\right]\right. \nonumber \\
&& \left.-\frac12\frac1{(1+\Xi_j)^2}\frac{\eta^2}{(\eta+1)^2}
\right\}~~;\qquad y^\dagger=1~~;
\label{seq:wiH}
\end{leftsubeqnarray}
\begin{leftsubeqnarray}
\slabel{eq:weHa}
&& w^{({\rm ext})}(\eta)=-128\left\{-\frac12\frac1{(y^\dagger-1)^4}
\left[\frac{(y^\dagger-1)^2\eta}{(\eta+1)^2}+\frac
{2(y^\dagger)^2(y^\dagger-1)\eta}{1+y^\dagger\eta}\right.\right. \nonumber \\
&& \left.+\frac{(y^\dagger-1)(3y^\dagger+1)\eta}{\eta+
1}-2y^\dagger(y^\dagger+2)\ln\frac{y^\dagger\eta+1}{\eta+1}\right] \nonumber \\
&& \left.-\frac12\frac1{(1+\Xi_i)^2}
\frac{\eta^2}{(\eta+1)^2}\right\}~~;\qquad y^\dagger\ne1~~; \\
\slabel{eq:weHb}
&& w^{({\rm ext})}(\eta)=128\left\{\frac1{12}\left[\frac{4\eta+1}
{(\eta+1)^4}-1\right]\right. \nonumber \\
&& \left.+\frac12\frac1{(1+\Xi_i)^2}\frac{\eta^2}{(\eta+1)^2}
\right\}~~;\qquad y^\dagger=1~~;
\label{seq:weH}
\end{leftsubeqnarray}
using Eqs.\,(\ref{eq:nuuvb}) and
(\ref{eq:numH})-(\ref{seq:weH}), the HH
macrogases equation of state is obtained
from the particularization of Eq.\,(\ref
{eq:phi}) to the case of interest for
the domain, $y\ge1$.
The extension
to the domain, $0\le y\le1$, can be
done following the procedure outlined
in Subsection \ref{pete}.

In absence of truncation radius, the density
drops to zero when the radius goes to
infinity, $\Xi\to+\infty$, $\eta\to+\infty$,
and Eqs.\,(\ref{eq:FH})-(\ref{seq:weH})
reduce to:
\begin{lefteqnarray}
\label{eq:FlH}
&& \lim_{\Xi_u\to+\infty}F_u(\xi_u)=\frac8{(1+\xi_u)^2}~~;\qquad u=i,j~~; \\
\label{eq:lumH}
&& \lim_{\Xi_u\to+\infty}(\nu_u)_{\rm mas}=12~~;\qquad u=i,j~~; \\
\label{eq:lusH}
&& \lim_{\Xi_u\to+\infty}(\nu_u)_{\rm sel}=12~~;\qquad u=i,j~~;
\end{lefteqnarray}
\begin{leftsubeqnarray}
\slabel{eq:wilHa}
&& \lim_{\eta\to+\infty}w^{({\rm int})}(\eta)=-\frac{64y^\dagger}
{(y^\dagger-1)^4}\left[-2(2y^\dagger+1)\ln y^\dagger+(y^\dagger-1)
(y^\dagger+5)\right]~~; \nonumber \\
&& y^\dagger\ne1~~; \\
\slabel{eq:wilHb}
&& \lim_{\eta\to+\infty}w^{({\rm int})}(\eta)=-\frac{32}3~~;\qquad
y^\dagger=1~~;
\label{seq:wilH}
\end{leftsubeqnarray}
\begin{leftsubeqnarray}
\slabel{eq:welHa}
&& \lim_{\eta\to+\infty}w^{({\rm ext})}(\eta)=-\frac{64}
{(y^\dagger-1)^4}\left[2y^\dagger(y^\dagger+2)\ln y^\dagger-(y^\dagger-1)
(5y^\dagger+1)\right]~~; \nonumber \\
&& y^\dagger\ne1~~; \\
\slabel{eq:welHb}
&& \lim_{\eta\to+\infty}w^{({\rm ext})}(\eta)=-\frac{32}3~~;\qquad
y^\dagger=1~~;
\label{seq:welH}
\end{leftsubeqnarray}
where, in particular, the related expression
of the tidal energy coincides with its
counterpart calculated in an earlier attempt
(Valentinuzzi, 2006, Chap.\,4, \S4.2.1).
Using Eqs.\,(\ref{eq:nuuvb}) and
(\ref{eq:lumH})-(\ref{seq:welH}), the HH
macrogases equation of state in the
special situation under discussion, is
obtained from the particularization of
Eq.\,(\ref{eq:phi}) to the case of interest,
for the domain, $y\ge0$.
If, in particular, $y=y^\dagger=1$,
$\Xi_j=\Xi_i$, the combination of
Eqs.\,(\ref{eq:phi}) and (\ref
{eq:lumH})-(\ref{seq:welH}) yields
Eq.\,(\ref{eq:ph1U}).

\subsection{HP macrogases}\label{HP}

The inner density profile (Hernquist,
1990) implies $(\alpha,\beta,\gamma)=
(1,4,1)$, which is defined by
Eq.\,(\ref{eq:fH}), and related
functions and parameters by
Eqs.\,(\ref{eq:FH})-(\ref{eq:nusH})
or (\ref{eq:FlH})-(\ref{eq:lusH})
in the special case of no truncation
radius.

The outer density profile (Schuster,
1883; Plummer, 1911) implies $(\alpha,
\beta,\gamma)=(2,5,0)$, which is defined
by Eq.\,(\ref{eq:fP}), and related
functions and parameters by
Eqs.\,(\ref{eq:FP})-(\ref{eq:nusP})
or (\ref{eq:FlP})-(\ref{eq:lusP})
in the special case of no truncation
radius.

With regard to the tidal potential-energy
terms, the particularization of
Eqs.\,(\ref{eq:wiea}) and (\ref{eq:wieb})
to the case of interest yields:
\begin{leftsubeqnarray}
\slabel{eq:wiHPa}
&& w^{({\rm int})}(\eta)=-\frac{128\sqrt{2}}3y^\dagger\left[\Phi_1^{({\rm
int})}(\eta)+\Phi_2^{({\rm int})}(\eta)-\frac12\frac{\eta^2}{(1+\Xi_j)^{3/2}}
\right]~~; \\
\slabel{eq:wiHPb}
&& \Phi_1^{({\rm int})}(\eta)=\frac{(y^\dagger)^4-12(y^\dagger)^2+2}{2[(y^
\dagger)^2+1]^3}+\frac{P^{({\rm int})}(\eta)}{2[(y^\dagger)^2+1]^3(y^\dagger
\eta+1)^2(\eta^2+1)^{1/2}}~~;\qquad \\
\slabel{eq:wiHPc}
&& \Phi_2^{({\rm int})}(\eta)=-\frac{3y^\dagger[3(y^\dagger)^2-2]}{2[(y^
\dagger)^2+1]^{7/2}}\ln\frac{\left[y^\dagger-\eta+\sqrt{(y^\dagger)^2+1}\sqrt
{\eta^2+1}\right]}{(y^\dagger\eta+1)\left[y^\dagger+\sqrt{(y^\dagger)^2+1}
\right]}~~; \\
\slabel{eq:wiHPd}
&& P^{({\rm int})}(\eta)=-2(y^\dagger)^5\eta(2\eta^2+1)+(y^\dagger)^4(\eta^2-
1)+(y^\dagger)^3\eta(11\eta^2+15) \nonumber \\
&& \phantom{P^{({\rm int})}(\eta)=}+4(y^\dagger)^2(4\eta^2+3)+2y^\dagger\eta-
2~~;
\label{seq:wiHP}
\end{leftsubeqnarray}
\begin{leftsubeqnarray}
\slabel{eq:weHPa}
&& w^{({\rm ext})}(\eta)=-64\sqrt{2}\left[\Phi_1^{\rm ext}
(\eta)+\Phi_2^{\rm ext}(\eta)-\frac13\frac1{(1+\Xi_i)^2}\frac{{\eta^3}}
{(1+\eta^2)^{3/2}}\right]~~; \\
\slabel{eq:weHPb}
&& \Phi_1^{\rm (ext)}(\eta)=\frac{11(y^\dagger)^3-4y^\dagger}{3[(y^
\dagger)^2+1]^3}+\frac{P^{({\rm ext})}(\eta)}{3[(y^\dagger)^2+1]^3(y^\dagger
\eta+1)(\eta^2+1)^{3/2}}~~; \\
\slabel{eq:weHPc}
&& \Phi_2^{\rm (ext)}(\eta)=\frac{(y^\dagger)^2[2(y^\dagger)^2-3]}{[(y^
\dagger)^2+1]^{7/2}}\ln\frac{\left[y^\dagger-\eta+\sqrt{(y^\dagger)^2+1}\sqrt
{\eta^2+1}\right]}{(y^\dagger\eta+1)\left[y^\dagger+\sqrt{(y^\dagger)^2+1}
\right]}~~; \\
\slabel{eq:weHPd}
&& P^{({\rm ext})}(\eta)=(y^\dagger)^5\eta^2(2\eta^2+3)-(y^\dagger)^4\eta
(4\eta^2+5)-(y^\dagger)^3(12\eta^4+21\eta^2+11) \nonumber \\
&& \phantom{P^{({\rm ext})}(\eta)=}-(y^\dagger)^2\eta(3\eta^2+5)+
y^\dagger(\eta^4+6\eta^2+4)+\eta^3~~;
\label{seq:weHP}
\end{leftsubeqnarray}
using Eqs.\,(\ref{eq:nuuvb}), (\ref{eq:numP}),
(\ref{eq:nusP}),
(\ref{eq:numH}), and (\ref{eq:nusH}), the HP
macrogases equation of state is obtained
from the particularization of Eq.\,(\ref
{eq:phi}) to the case of interest for
the domain, $y\ge1$.
%The extension
%to the domain, $0\le y\le1$, can be
%done following the procedure outlined
%in subsection \ref{pete}.

In absence of truncation radius, the density
drops to zero when the radius goes to
infinity, $\Xi\to+\infty$, $\eta\to+\infty$,
and Eqs.\,(\ref{eq:FP})-(\ref{eq:nusP}),
(\ref{eq:FH})-(\ref{eq:nusH}), reduce to
(\ref{eq:FlP})-(\ref{eq:lusP}),
(\ref{eq:FlH})-(\ref{eq:lusH}), respectively,
and Eqs.\,(\ref{seq:wiHP})-(\ref{seq:weHP})
reduce to:
\begin{leftsubeqnarray}
\slabel{eq:liHPa}
&& \lim_{\eta\to+\infty}w^{({\rm int})}(\eta)=-\frac{64\sqrt{2}y^\dagger}
{3[(y^\dagger)^2+1]^3}\left\{[(y^\dagger)^4-4(y^\dagger)^3-
12(y^\dagger)^2+11y^\dagger+2]
\phantom{\frac{\sqrt{(y^\dagger)^2}}{\sqrt{(y^\dagger)^2}}}
\right. \nonumber \\
&& \left.-\frac{3y^\dagger[3(y^\dagger)^2-2]}{\sqrt{(y^\dagger)^2+1}}\ln\frac
{\sqrt{(y^\dagger)^2+1}-1}{y^\dagger\left[y^\dagger+\sqrt{(y^\dagger)^2+1}
\right]}\right\}~~; \qquad y^\dagger\ne1~~; \\
\slabel{eq:liHPb}
&& \lim_{\eta\to+\infty}w^{({\rm int})}(\eta)=\frac83\left[2\sqrt2+3\ln\frac
{\sqrt2-1}{\sqrt2+1}\right]~~;\qquad y^\dagger=1~~;
\label{seq:liHP}
\end{leftsubeqnarray}
\begin{leftsubeqnarray}
\slabel{eq:leHPa}
&& \lim_{\eta\to+\infty}w^{({\rm ext})}(\eta)=-\frac{64\sqrt{2}}
{3[(y^\dagger)^2+1]^3}\left\{[2(y^\dagger)^4+11(y^\dagger)^3-12(y^\dagger)^2-4
y^\dagger+1]
\phantom{\overline{\frac{\overline{\sqrt{y^2}}}{\overline{\sqrt{y^2}}}}}
\right. \nonumber \\
&& \left.-\frac{3(y^\dagger)^2[-2(y^\dagger)^2+3]}{\sqrt{(y^\dagger)^2+1}}\ln
\frac{\sqrt{(y^\dagger)^2+1}-1}{y^\dagger\left[y^\dagger+\sqrt{(y^\dagger)^2+
1}\right]}\right\}\qquad y^\dagger\ne1~~; \\
\slabel{eq:leHPb}
&& \lim_{\eta\to+\infty}w^{({\rm ext})}(\eta)=\frac83\left[2\sqrt2+3\ln\frac
{\sqrt2-1}{\sqrt2+1}\right]~~;\qquad y^\dagger=1~~;
\label{seq:leHP}
\end{leftsubeqnarray}
where, in particular, the related expression
of the tidal energy coincides with its
counterpart calculated in an earlier attempt
(Valentinuzzi, 2006, Chap.\,4, \S4.2.2).
Using Eqs.\,(\ref{eq:nuuvb}) and
(\ref{eq:lumP}), (\ref{eq:lusP}),
(\ref{eq:lumH}), (\ref{eq:lusH}),
(\ref{seq:wiHP}), (\ref{seq:weHP}), the HP
macrogases equation of state in the
special situation under discussion, is
obtained from the particularization of
Eq.\,(\ref{eq:phi}) to the case of interest,
for the domain, $y\ge0$.

\subsection{HN macrogases}\label{HN}

The inner density profile (Hernquist,
1990) implies $(\alpha,\beta,\gamma)=
(1,4,1)$, which is defined by
Eq.\,(\ref{eq:fH}), and related
functions and parameters by
Eqs.\,(\ref{eq:FH})-(\ref{eq:nusH})
or (\ref{eq:FlH})-(\ref{eq:lusH})
in the special case of no truncation
radius.

The outer density profile (Navarro et
al., 1995, 1996, 1997) implies $(\alpha,
\beta,\gamma)=(1,3,1)$,
and Eq.\,(\ref{eq:ex3}) reduces to:
\begin{equation}
\label{eq:fN}
f_u(\xi_j)=\frac4{\xi_j(1+\xi_j)^2}~~;
\qquad0\le\xi_j\le\Xi_j~~;
\end{equation}
which, together with its H counterpart
expressed by Eq.\,(\ref{eq:fH}),
have been proved to be consistent with
nonnegative distribution functions,
in the parameter range of interest
(Lowenstein and White, 1999).

The particularization of Eqs.\,(\ref
{eq:Fa}), (\ref{eq:Mb}), and (\ref{eq:Esc})
to the case of interest yields:
\begin{lefteqnarray}
\label{eq:FN}
&& F_j(\xi_j)=\frac8{1+\xi_j}-\frac8{1+\Xi_j}~~; \\
\label{eq:numN}
&& (\nu_j)_{\rm mas}=12\left[\ln(1+\Xi_j)-\frac{\Xi_j}{1+\Xi_j}\right]~~; \\
\label{eq:nusN}
&& (\nu_j)_{\rm sel}=36\frac{\Xi_j(2+\Xi_j)-2(1+\Xi_j)\ln(1+\Xi_j)}
{(1+\Xi_j)^2}~~;
\end{lefteqnarray}
with regard to the tidal potential-energy
terms, the particularization of
Eqs.\,(\ref{eq:wiea}) and (\ref{eq:wieb})
to the case of interest yields:
\begin{leftsubeqnarray}
\slabel{eq:wiHNa}
&& w^{({\rm int})}(\eta)=-\frac{64y^\dagger}{(y^\dagger-1)^3}\left[-\frac
{(y^\dagger-1)^2}{y^\dagger}\frac{y^\dagger\eta(y^\dagger\eta+2)}
{(y^\dagger\eta+1)^2}+\frac{2y^\dagger\eta(y^\dagger-1)}{y^\dagger\eta+1}
\right. \nonumber \\
&& \left.+2\ln\frac{\eta+1}{y^\dagger\eta+1}-\frac{(y^\dagger-1)^3}{1+\Xi_j}
\frac{\eta^2}{(y^\dagger\eta+1)^2}\right]~~;\qquad y^\dagger\ne1~~; \\
\slabel{eq:wiHNb}
&& w^{({\rm int})}(\eta)=-\frac{64\eta^2}{(\eta+1)^2}\left[\frac{\eta+3}
{3(\eta+1)}-\frac1{1+\Xi_j}\right]~~;\qquad y^\dagger=1~~;
\label{seq:wiHN}
\end{leftsubeqnarray}
\begin{leftsubeqnarray}
\slabel{eq:weHNa}
&& w^{({\rm ext})}(\eta)=-\frac{64}{(y^\dagger-1)^2}\left\{-\frac\eta{\eta+1}-
\frac{y^\dagger\eta}{y^\dagger\eta+1}-\frac{y^\dagger+1}{y^\dagger-1}\ln\frac
{\eta+1}{y^\dagger\eta+1}\right. \nonumber \\
&& \left.-\frac{(y^\dagger-1)^2}{(1+\Xi_i)^2}\left[-\frac\eta{\eta+1}+\ln
(\eta+1)\right]\right\}~~;\qquad y^\dagger\ne1~~; \\
\slabel{eq:weHNb}
&& w^{({\rm ext})}(\eta)=-64\left\{\frac16\frac{\eta^2(\eta+3)}{(\eta+1)^3}
-\frac1{(1+\Xi_i)^2}\left[-\frac\eta{\eta+1}+\ln(\eta+1)\right]\right\}~~;
\nonumber \\
&& y^\dagger=1~~;
\label{seq:weHN}
\end{leftsubeqnarray}
using Eqs.\,(\ref{eq:nuuvb}), (\ref{eq:numH}),
(\ref{eq:nusH}),
(\ref{eq:numN}), and (\ref{eq:nusN}), the HN
macrogases equation of state is obtained
from the particularization of Eq.\,(\ref
{eq:phi}) to the case of interest for
the domain, $y\ge1$.
%The extension
%to the domain, $0\le y\le1$, can be
%done following the procedure outlined
%in subsection \ref{pete}.

In absence of truncation radius, the density
drops to zero when the radius goes to
infinity, $\Xi\to+\infty$, $\eta\to+\infty$,
and Eqs.\,(\ref{eq:FH})-(\ref{eq:nusH}),
reduce to (\ref{eq:FlH})-(\ref{eq:lusH}),
respectively, and
Eqs.\,(\ref{eq:FN})-(\ref{seq:weHN})
reduce to:
\begin{lefteqnarray}
\label{eq:FlN}
&& \lim_{\Xi_j\to+\infty}F_j(\xi_j)=\frac8{1+\xi_j}~~; \\
\label{eq:lumN}
&& \lim_{\Xi_j\to+\infty}(\nu_j)_{\rm mas}=+\infty~~; \\
\label{eq:lusN}
&& \lim_{\Xi_j\to+\infty}(\nu_j)_{\rm sel}=36~~;
\end{lefteqnarray}
\begin{leftsubeqnarray}
\slabel{eq:liHNa}
&& \lim_{\eta\to+\infty}w^{({\rm int})}(\eta)=-\frac{64}
{(y^\dagger-1)^3}\left[\frac{(y^\dagger)^2-1}{y^\dagger}-2y^\dagger\ln
y^\dagger\right]~~;\qquad y^\dagger\ne1~~; \\
\slabel{eq:liHNb}
&& \lim_{\eta\to+\infty}w^{({\rm int})}(\eta)=-\frac{64}3~~;\qquad
y^\dagger=1~~;
\label{seq:liHN}
\end{leftsubeqnarray}
\begin{leftsubeqnarray}
\slabel{eq:leHNa}
&& \lim_{\eta\to+\infty}w^{({\rm ext})}(\eta)=-\frac{64}
{(y^\dagger-1)^2}\left[2-\frac{y^\dagger+1}{y^\dagger-1}\ln y^\dagger
\right]~~;\qquad y^\dagger\ne1~~; \\
\slabel{eq:leHNb}
&& \lim_{\eta\to+\infty}w^{({\rm ext})}(\eta)=-\frac{32}3~~;\qquad
y^\dagger=1~~;
\label{seq:leHN}
\end{leftsubeqnarray}
where the self potential-energy profile
factor remains finite, although the mass
profile factor undergoes a
logarithmic divergence.
Using Eqs.\,(\ref{eq:nuuvb}) and
(\ref{eq:lusN})-(\ref{seq:weHN}),
the HN macrogases equation of state
(where $M_j^\dagger$ and $m^\dagger$
appear instead of $M_j$ and $m$) in
the situation under consideration, is
obtained from the particularization of
Eq.\,(\ref{eq:phi}) to the case of interest,
for the domain, $y\ge0$.

\section{Results}
\label{resu}

The macrogases equation of state, Eq.\,(\ref
{eq:phi}), is represented in the next figures
for a number of cases analysed in Section
\ref{spec}.   In particular, owing to
Eqs.\,(\ref{eq:phlU}) and (\ref{eq:phuU}),
the UU macrogases equation of state is
independent of the scaled truncation
radii, $(\Xi_i,\Xi_j)$.

The macrogases equation of state in absence
of truncation radius, is plotted in
\begin{figure*}[t]
\begin{center}
\includegraphics[scale=0.8]{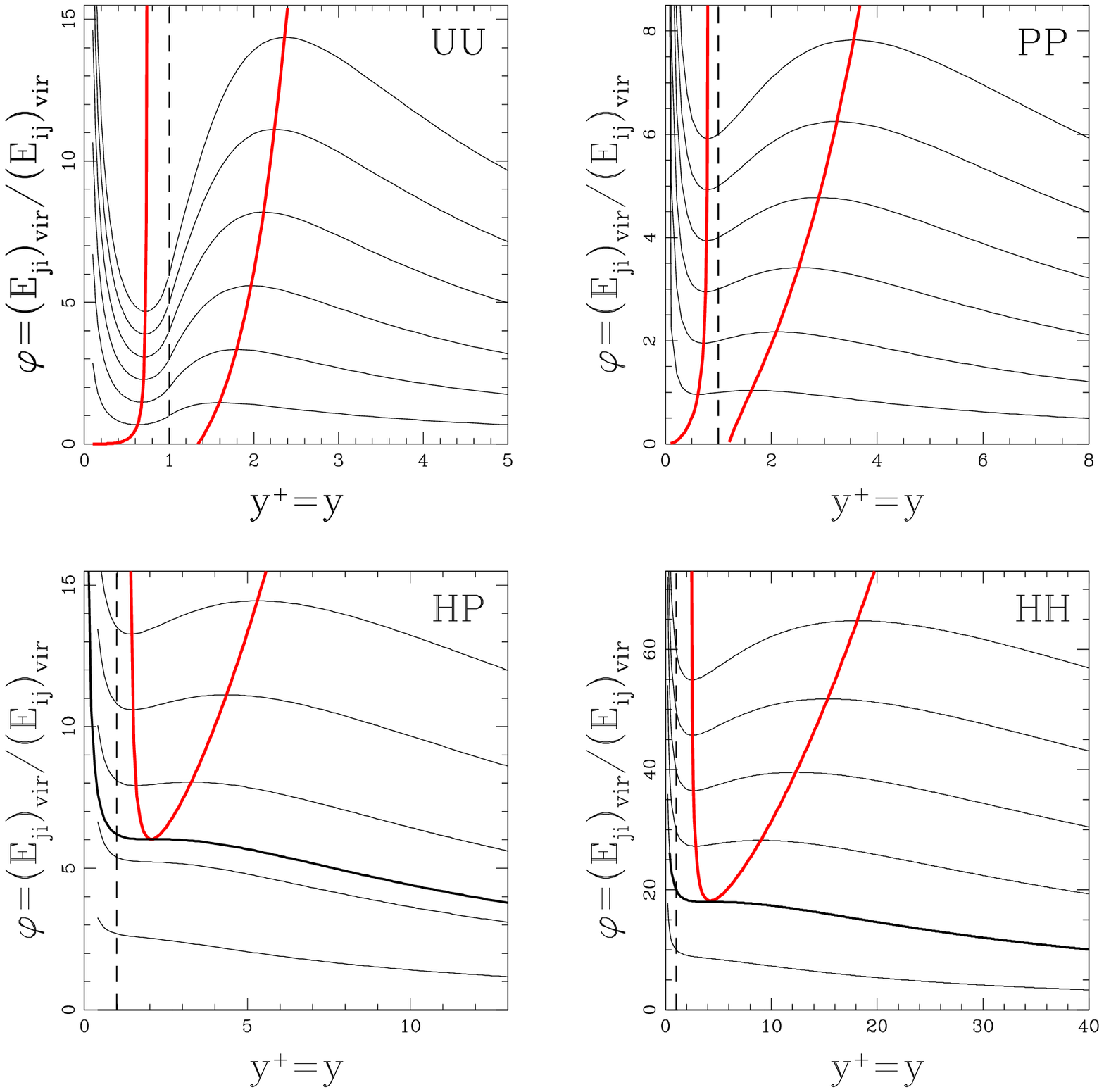}
\caption{Isofractional mass curves related to
(from top left in clockwise sense) UU, PP, HH,
HP, macrogases for $\Xi\to+\infty$,
$\Xi_j/\Xi_i\to1$.   Thick curves
%(red in the online version of the paper)
represent
the locus of minimum (left branch) and maximum
(right branch) points.   The critical curve
is also thickened, and the horizontal inflexion
point, or critical point, is defined (when
present) as the tangent point with respect
to the locus of the extremum points.   The
value of the fractional mass
is $m=\phi(1,m)$ and can be read on the
intersection between the selected curve
and the dashed vertical line, $y^\dagger=y=
1$, provided the density profile of the inner
and the outer subsystem belong to the same
family (UU, PP, HH).   All the curves diverge
at $y\to0$ and converge to 0 at $y\to+\infty$.
For UU macrogases, the extremum points of the
$m=0$ isofractional mass curve occur at $y=0
$ and $y=3/\sqrt{5}$, respectively, and the
locus of the minimum points has a vertical
asymptote, $y=\sqrt{5}/3$.   For PP macrogases,
the extremum points of the $m=0$ isofractional
mass curve occur at $y=0$ and $y\approx1.20$,
respectively, and the locus of the minimum
points has a vertical asymptote, $y\approx0.84$.}
\label{f:inf}
\end{center}
\end{figure*}
Fig.\,\ref{f:inf} for cases (from top left in
clockwise sense) UU, PP, HH, HP, where $\Xi_j=
\Xi_i$ and (from bottom to top in each panel)
$m=$1, 2, ..., 6, for cases UU, PP, HP, and
$m=$10, 20, ..., 60, for case HH.
Shallower density profiles (UU, PP) show the
occurrence of two extremum points: one
minimum on the left and one maximum on the
right, for arbitrary values of the fractional
mass, $m$.
%The abscissa of the minima is less than unity.
Steeper density profiles (HP,
HH) still exhibit extremum points,
but none below a treshold, where the critical
isofractional mass curve shows a single horizontal
inflexion point.
%The abscissa of the minima is more than unity.

As already noticed in an
earlier attempt (CS90), the same trend is
shown by van der Waals' (1873) isothermal
curves and, in fact, the van der Waals'
equation of state looks similar to the
macrogases equation of state, Eq.\,(\ref
{eq:phi}), where three variables also appear.
Thick curves
%(red in the online version of the paper)
represent the locus of
minimum (left branch) and maximum (right
branch) points. The critical isofractional
mass curve, when it appears, is also thickened.

A main feature is that, above the critical
isofractional mass curve, for any selected
value of $m$, a range in fractional virial
potential energy,
$\phi$, exists, $\phi_{\rm min}<\phi<
\phi_{\rm max}$, where three different
configurations i.e. different fractional
truncation radius, $y$, correspond to the same value
of $\phi$.   By analogy with van der
Waals' isothermal curves, it could be
argued the existence of a bell-shaped
region in the ${\sf O}y\phi$ plane,
within which the extremum points are
located, and where a phase transition
occurs.   Further investigation is
needed on this point, but it lies outside
the aim of the current paper.

\begin{figure*}[t]
\begin{center}
\includegraphics[scale=0.8]{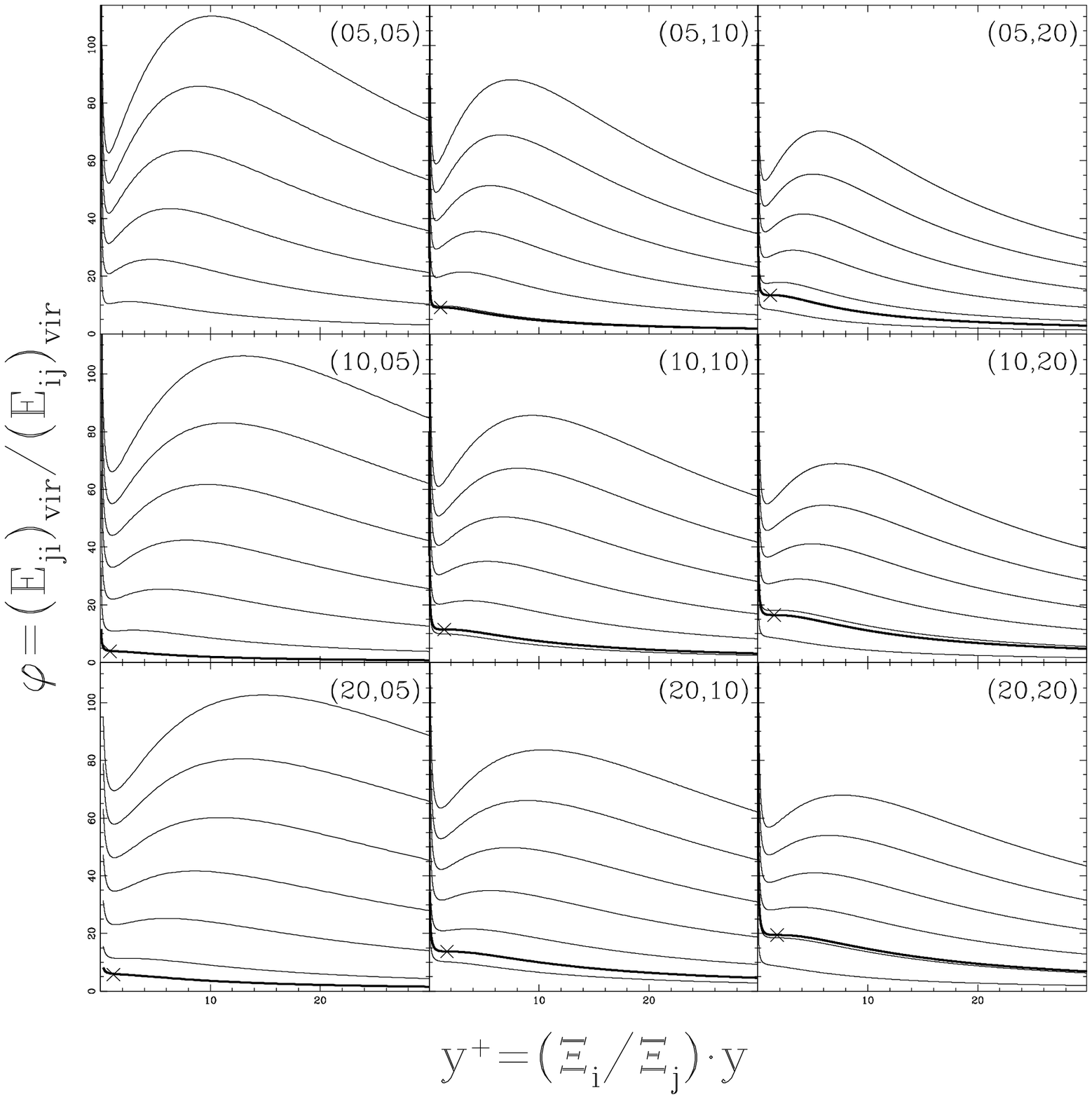}
\caption{Isofractional mass curves related
to HH macrogases, for different choices of
scaled truncation radii, $(\Xi_i,
\Xi_j)$, represented on each panel, where
(from bottom to top) $m=$10, 20, ..., 60.
The critical isofractional mass
curve, where the horizontal
inflexion point, or critical point, is
marked by a St.\,Andrew's cross, is
thickened on each
panel.   All the curves diverge at $y\to0$
and converge to 0 at $y\to+\infty$.}
\label{f:HH}
\end{center}
\end{figure*}

\begin{figure*}[t]
\begin{center}
\includegraphics[scale=0.8]{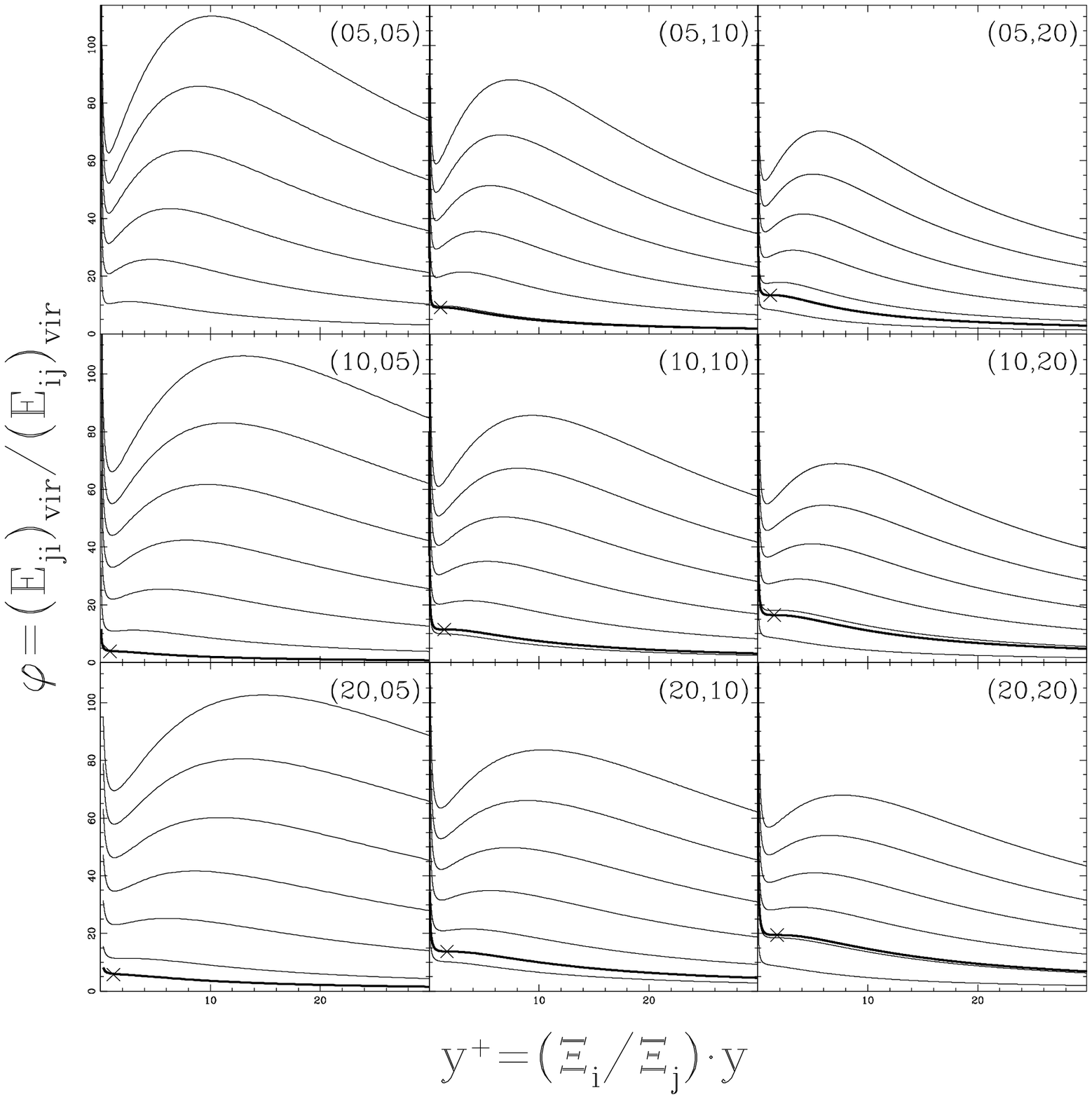}
\caption{Isofractional mass curves related
to HN macrogases, for different choices of
scaled truncation radii, $(\Xi_i,
\Xi_j)$, represented on each panel, where
(from bottom to top) $m=$10, 20, ..., 60.
The critical isofractional mass
curve, where the horizontal
inflexion point, or critical point, is
marked by a St.\,Andrew's cross, is
thickened on each
panel.   All the curves diverge at $y\to0$
and converge to 0 at $y\to+\infty$.}
\label{f:HN}
\end{center}
\end{figure*}

The macrogases equation of state in
presence of truncation radius, is
plotted in Figs.\,\ref{f:HH} and
\ref{f:HN}, respectively, for
different choices of the fractional
truncation radii, $(\Xi_i, \Xi_j)$,
represented on each panel where (from
bottom to top) $m=$10, 20, ..., 60,
and approximate values of the
parameters related to the
critical point i.e. the horizontal
inflexion point on the critical
isofractional mass curve (marked
by a St.\,Andrew's cross), are listed
in Tab.\,\ref{t:crit}.

It can be
\begin{table}
\caption{Approximate values of parameters
related to the
critical point i.e. the horizontal inflexion
point on the critical isofractional mass
curve, for the density profiles under
investigation.   Bars instead of numbers
mean absence of critical curve in the
corresponding case.   Numbers in brackets
denote the values of scaled truncation
radii, $(\Xi_i, \Xi_j)$.   In absence of
truncation radius, $\Xi\to+\infty$, the
case considered is $\Xi_j/\Xi_i\to1$,
which implies $y^\dagger=y$.   For
homogeneous configurations (case UU)
the isofractional mass curves are
independent of $(\Xi_i, \Xi_j)$.}
\label{t:crit}
\begin{center}
\begin{tabular}{|c|c|c|c|c|c|c|} \hline
case &   ($\Xi_i$,$\Xi_j$)& $m^\dagger$ & $m$ & $y^\dagger$ & $y$ &
$\varphi$ \\
\hline
HN & (05-05)             & ---   & ---   & ---  & ---   & ---  \\
   & (05-10)             & 04.41 & 09.45 & 1.04 & 2.08 & 09.16 \\
   & (05-20)             & 05.14 & 15.49 & 1.14 & 4.56 & 13.44 \\
   & (10-05)             & 03.17 & 03.68 & 0.88 & 0.44 & 03.90 \\
   & (10-10)             & 06.36 & 11.46 & 1.38 & 1.38 & 11.48 \\
   & (10-20)             & 07.24 & 18.33 & 1.50 & 3.00 & 16.47 \\
   & (20-05)             & 04.99 & 05.27 & 1.18 & 0.29 & 05.75 \\
   & (20-10)             & 08.10 & 13.30 & 1.62 & 0.81 & 13.74 \\
   & (20-20)             & 09.10 & 20.99 & 1.75 & 1.75 & 19.48 \\
\hline
HH & (05-05)             & 07.10 & 07.10 & 2.32 & 2.32 & 06.86 \\
   & (05-10)             & 07.72 & 09.19 & 2.43 & 4.86 & 08.17 \\
   & (05-20)             & 07.90 & 10.32 & 2.45 & 9.80 & 08.62 \\
   & (10-05)             & 11.15 & 09.37 & 3.18 & 1.59 & 09.44 \\
   & (10-10)             & 11.88 & 11.88 & 3.30 & 3.30 & 11.03 \\
   & (10-20)             & 12.10 & 13.28 & 3.33 & 6.66 & 11.57 \\
   & (20-05)             & 15.00 & 11.50 & 3.77 & 0.94 & 12.04 \\
   & (20-10)             & 15.83 & 14.42 & 3.88 & 1.94 & 13.94 \\
   & (20-20)             & 16.08 & 16.08 & 3.92 & 3.92 & 14.60 \\
   & ($\infty$-$\infty$) & 20.22 & 20.22 & 4.27 & 4.27 & 18.15 \\
\hline
HP & ($\infty$-$\infty$) & 04.59 & 04.59 & 2.05 & 2.05 & 06.03 \\
\hline
PP & ($\infty$-$\infty$) & ---   & ---   & ---  & ---  & ---   \\
\hline
UU & ($\Xi_i$,$\Xi_j$) & ---   & ---   & ---  & ---  & ---   \\
\hline
\end{tabular}
\end{center}
\end{table}
seen that the occurrence of the truncation
radius makes little change to the trend of
the isofractional mass curves.   In general,
decreasing the outer scaled truncation
radius, $\Xi_j$, yields more pronounced
extremum points and vice versa.   Much smaller
changes, in the same sense, are produced by
decreasing the inner scaled truncation
radius, $\Xi_i$.   This is why, in the cases
under discussion, the mass of the outer
subsystem is dominant with respect to the
mass of the inner subsystem, $m\gg1$.
The critical isofractional mass curve
does not appear for earlier truncated
HN density profiles, $(\Xi_i, \Xi_j)=
(5,5)$, which are shallower in comparison
with their latter truncated counterparts.
%As in absence of truncation radius, the
%abscissa of the minima is less than
%unity for shallower density profiles,
%where the critical isofractional mass
%curve is absent, and more than unity
%for steeper density profiles, where
%the critical isofractional mass curve
%is present.

Reduced truncation radii, $\Xi=
R/r^\dagger$, may be conceived as
concentrations of the related matter
distribution (e.g., Navarro et al.,
1997; CM03; Caimmi et al., 2005) and
the selected range, $5\le\Xi\le20$,
is consistent with the results of
dark matter (hereafter quoted as DM)
halo numerical simulations (e.g., Bullock
et al., 2001) and elliptical galaxy
(hereafter quoted as EG) observations
(e.g., Lowenstein and White, 1999).

\section{Application to elliptical galaxies
and their hosting haloes}
\label{appl}

According to current cosmological scenarios
(e.g., Mota and van de Bruck, 2004; Percival,
2005; Horellou and Berge, 2005; Maor and Lahav,
2005; Nunes and Mota, 2006), density perturbations
at recombination epoch ($z\approx1100$) initially
expand with the universe, turn around, collapse,
and finally virialize (at least in their inner and
denser regions).
%Just after virialization has been
%attained, baryonic and (dissipationless) non baryonic
%DM are expected to fill the same volume and
%to be described by the same density profile.
%At late times the situation changes, as energy
%dissipation within the gaseous subsystem
%causes further contraction, whereas the
%DM does not appreciably change in extension.

Virialized density perturbations, such as
typical elliptical galaxies (EGs)
and clusters of galaxies, may be
idealized, to a first extent, as two
homeoidally striated, similar and
similarly placed, density profiles.
In the following, attention
shall be focused on EGs.

\subsection{General considerations and main assumptions}
\label{geco}

A recent investigation performed on an optically
complete sample of 42 EGs,
for which X-ray gas temperatures and central
stellar velocity dispersions were determined
(Davis \& White 1996), has shown evidence that,
in general, EGs contain substantial
amounts of DM (Loewenstein and White, 1999).
Accordingly, more than about
20\% and 39\%-85\% of the total mass
within one and six optical radii, respectively,
is in form of (non baryonic) DM,
depending on the stellar density profile and
observed value of X-ray gas temperature and
central stellar velocity dispersion.   The
comparison between the velocity dispersion
distributions for DM and
stars, assuming isotropic orbits, shows
that the DM is dynamically
``hotter'' than the stars, by a factor
1.4-2 (Loewenstein and White, 1999).

The above investigation cannot be considered
as conclusive in favour of the existence of
DM haloes hosting EGs.
In fact, it has been pointed out
that the attenuation (in particular, the
scattering) by dust grains has the same effect
on the stellar kinematics as a DM
halo (Baes and Dejonghe, 2001).
According to a recent attempt,
no strong evidence for DM haloes within
1-3 optical radii has been found in a
restricted sample of 4 early-type
EGs, using dynamical modelling (Samurovic
and Danziger, 1995).
Beyond 1-3 optical radii, the X-ray methodology
shows the need of DM where an X-ray
halo is detected (Samurovic and Danziger,
1995).   A similar trend is exhibited by an
additional early-type EG (Samurovic and
Danziger, 1996).    In any case,
current cosmological scenarios ($\Lambda$CDM,
QCDM) predict DM haloes hosting
EGs, as well as spiral galaxies, for which
there are lots of
empirical evidence (e.g., flat rotation curves
well outside optical radii).   For this
reason, EGs are also
assumed to be embedded within DM
haloes.

An analysis on the evolution of the
physical properties of cosmological
baryons at low redshifts ($z\appleq
5$) has recently been performed
(Valageas et al., 2002), which (i)
yields robust model-independent
results that agree with numerical
simulations; (ii) recovers the fraction
of matter within different phases and
the spatial clustering computed by
numerical simulations; (iii) predicts
a soft X-ray background due to the
``warm'' intergalactic medium component,
that is consistent with observations.
The related baryon fraction in the
present universe is found to be
7\% in hot gas, 24\% in the warm
intergalactic medium, 38\% in the
cool intergalactic medium, 9\% within
star-like objects and, as a still
unobserved component, 22\% of dark
baryons associated with collapsed
structures, with a relative uncertainty
no larger than 30\% on these numbers.
Then the amount of still undetected
baryons is about one fifth of the total,
one fourth of the observed baryons
(intergalactic medium, stellar components,
and hot gas), and at least twice the
stellar-like component.

%EGs embedded within DM haloes, may safely be
%idealized as two homeoidally striated,
%similar and similarly placed, density
%profiles.
According to recent
investigations, HH (e.g., Ciotti, 1996)
or HN (e.g., Loewenstein and White, 1999)
models provide viable representations for the inner,
baryonic, and the outer, nonbaryonic subsystem,
respectively.
The above mentioned mass distributions were
found to be self-consistent, in the parameter
range of interest, with regard to the non
negativity of the distribution function
(e.g., Ciotti, 1996; Loewenstein and White,
1999) by use of a
theorem stated in an earlier attempt (Ciotti
and Pellegrini, 1992).

If undetected baryons in EGs
are present as hot gas, the gaseous
subsystem is expected to be less concentrated
than the stellar one, as in the Coma cluster
of galaxies (e.g., Briel et al. 1992).  If
undetected baryons in EGs are present as
unseen stars, the undetected subsystem is
expected to be as concentrated as the
stellar one.   Either assumption is
necessary in dealing with two-component
systems.   The real situation lies
between the above mentioned limiting
cases.

The typical velocity dispersion components,
deduced by use of the virial theorem
(hereafter referred to, in general,
as the virial velocity dispersions)
are global quantities, related
to the virial potential energy of the
subsystem as a whole, and so, by construction,
independent of the specific orbital distribution
of the particles.   This important property,
however, is also a weakness of the virial
theorem, in dealing with
velocity dispersion components
measured in the central region of a galaxy.
In fact, it is well known that the related
values can be significantly different for
structurally identical subsystems (and so
characterized by identical virial velocity
dispersion components), due to different
orbital structures (e.g., de Zeeuw \&
Franx 1991).   When using central
velocity dispersion components, an approach
based on Jeans equations (even though still
questionable) is to be preferred (e.g.,
Ciotti et al., 1996;
Ciotti \& Lanzoni 1997; Loewenstein and
White, 1999).   On the other hand, a comparison
between the results obtained by use of either
above mentioned methods, may provide additional
support to both of them and/or useful
indications on the nature of the problem
under investigation.

Strictly speaking, the central velocity
dispersions (along the line of sight) in
EGs, which are deduced
from observations, should be scaled to
the virial velocity dispersions.
Both observational evidence (e.g.,
Gerhard et al. 2001) and theoretical
arguments (e.g., Nipoti et al. 2002)
point towards the existence of dynamical
homology in EGs.   In
particular, a linear relation is found
between a local parameter, averaged
central velocity dispersion, and a
global parameter, inferred maximum
circular velocity, $\sigma_{0.1}=(2/3)
(v_c)_{max}$ (Gerhard et al. 2001).
Accordingly, the central velocity
dispersion components are expected
to be proportional to the virial
velocity dispersion components.
Then it could be assumed that the
related proportionality factor is of the
order of unity.
%, i.e. central velocity
%dispersion components are of the same
%order as corresponding virial velocity
%dispersion components.

In fact, typical peculiar velocity
component distributions within
EGs show a maximum
which is rapidly attained in the central
region (at about 1 kpc), and a slow
decrease occurs moving outwards (no
more than about 13\% the maximum at
about 10 kpc), at least in the case
of isotropic orbits; for further
details see related attempts (e.g.,
Loewenstein and White, 1999).
Accordingly, both the central and the
virial velocity dispersion components
are expected to be of comparable order,
slightly less than the maximum of the
peculiar velocity component distribution
(e.g., Cappellari et al., 2006).
On the other hand, most EGs
are moderately radially anisotropic (e.g.,
Gerhard et al. 2001), and the related
variation in central velocity dispersion
(an increase for increasing $\sigma_i^2$
and vice versa) is also expected to be
moderate.

%The above procedure has been followed in
%an earlier attempt (CM03).   For the
%current investigation, a method based
%on the Jeans equations (Ciotti et al.,
%1996) shall be used.

\subsection{Input parameters, specific
assumptions, and results}
\label{inpa}

The main assumptions of the
current model are (i) homeoidally
striated density profiles and (ii)
similar and similarly placed
boundaries.   Aspherical,
heterogeneuos, self-gravitating
fluids in dynamical or hydrostatic
equilibrium exhibit isopycnic
surfaces different from ellipsoids
(e.g., Chandrasekhar, 1933;
Chandrasekhar and Lebovitz, 1962;
Vandervoort and Welty, 1981; Lai
et al., 1993), and the above
assumption (i) is due to reasons
of simplicity.

Axisymmetric configurations with
nonsimilar boundaries have been
investigated in the special case
of homogeneous density profiles
i.e. UU macrogases, where the
shape of one component is kept
fixed and an additional variable,
the axis ratio of the other
component, together with an
additional relation, the angular
momentum conservation of the
other component, must be considered
(CS90).   A main feature is that
the isofractional mass curves on
the $({\sf O}y\phi)$ plane cannot
converge to 0 at $y\to+\infty$,
as the ending point occurs when
the inner subsystem attains a
flat configuration.   For further
details refer to the parent paper
(CS90).   Being the calculations
for heterogeneous density profiles
much more complicated, the above
assumption (ii) is also due to
reasons of simplicity.   It can
be expected that the effect due
to nonsimilar boundaries is
maximum for homogeneous density
profiles, and decreases as the
profile is steeper, to be null
for mass points surrounded by a
massless atmosphere (Roche
ellipsoids).   Then the results
found for homogeneous density
profiles (CS90) make a valid
reference for inhomogeneous
density profiles.
In conclusion, the above assumptions
(i) and (ii) are related to a viable
model which can be used for specific
applications to large-scale celestial
objects, in particular EGs.

Given a typical EG,
a natural question (in the
light of the model under discussion)
arises about its position on the
$({\sf O}y\phi)$ plane for assigned
density profiles and specified input
parameters.   Towards this aim, the
following main assumptions are made: (a) the
stellar and the DM distributions
are described by homeoidally striated,
similar and similarly placed, HH or HN
density profiles, respectively; (b)
undetected baryons trace either DM
haloes or EGs; and (c) the virial theorem
holds for each subsystem within the
tidal potential of the other one.

According to the above assumptions, a
typical EG embedded in
a DM halo is idealized
as two homeoidally striated, similar and
similarly placed matter distributions,
where the star and non baryonic subsystem
are described by HH or HN density
profiles, respectively.
%From this point
%on, the inner and the outer subsystem
%shall be labelled as $\ast$, $D$, instead
%of $i$, $j$, respectively.   Let us suppose,
%at the moment, that undetected baryons
%are absent.

For assigned density profiles, the
macrogases equation of state,
Eq.\,(\ref{eq:phi}), depends on two
independent variables: the fractional
truncation radius, $y=R_j/R_i$, and the fractional
mass, $m=M_j/M_i$, or their scaling
counterparts, $y^\dagger=r_j^\dagger/
r_i^\dagger$ and $m^\dagger=M_j^\dagger/
M_i^\dagger$.   To represent celestial
objects in the $({\sf O}y\phi)$ plane,
two additional equations are requested.
Having in mind an application to EGs
and their hosting haloes,
denoted in the following by the index
$i$ and $j$, respectively, for $y\ge1$,
an additional relation between the
stellar projected
velocity dispersion, $(\sigma_i)_{R_e/8}$,
averaged over the aperture  used for
spectroscopic observations $(r=R_e/8)$, and
the fractional scaling radius, $y^
\dagger$, may be deduced from the
virial theorem related to the star
subsystem, Eq.\,(\ref{eq:vir1b}),
by solving the appropriate Jeans
equations, following an earlier
approach (Ciotti et al., 1996).
The result may be written as:
\begin{equation}
\label{eq:sig0i}
(\sigma_i)_{R_e/8}^2=\frac{GM_i}{r_i^\dagger}\psi_i(y^\dagger,m^\dagger)~~;
\end{equation}
where $\psi_i$ is a function calculated by
a numerical algorithm.

Using Eq.\,(\ref{eq:sig0i}) implies
further restrictions, as its validity
is limited to (i) spherical-symmetric
matter distributions; (ii) isotropic
peculiar velocity distributions; (iii)
infinite truncation radii; (iv) HH
density profiles.   Concerning points
(i)-(iii) mentioned above, acceptable
approximations may safely be expected.
On the other hand, HH density profiles
provide a viable description to EGs
embedded within DM haloes (e.g.,
Ciotti, 1996).

At this stage, an additional relation
is needed.
The mere existence of a fundamental plane
(Djorgovski and Davis, 1987; Dressler et
al., 1987) indicates that structural
properties in EGs span a narrow range,
suggesting that some self-regulating
mechanism must be at work during formation
and evolution.   In particular, projected
light profiles from EGs exhibit large degree
of homogeneity and may well be fitted by the
$r^{1/4}$ de Vaucouleurs law.   Accordingly,
a narrow range may safely be expected
also for EG fractional mass and the
assumption, $m={\rm const}$, appears to
be a viable approximation.   The last,
together with Eqs.\,(\ref{eq:phi}) and
(\ref{eq:sig0i}), makes a system of three
equations in the three unknowns, $\phi$,
$y$, $m$, via Eqs.\,(\ref{seq:ym}), for
HH density profiles.   Then
the position of EGs
and their hosting haloes on a
selected isofractional mass curve in the
$({\sf O}y\phi)$ plane, can be determined.

The sample used $(N=16)$ is extracted
from a larger sample $(N=25)$ of EGs
and lenticular galaxies investigated
within the SAURON project (Cappellari et
al., 2006, Tab.\,1 therein), for which
the parameters of
interest i.e. masses, projected central
velocity dispersions, and effective
radii, can be derived from the data
listed in Tab.\,\ref{t:sauron}.
\begin{table}
\caption[]{Data
related to a subsample $(N=16)$ extracted
from a sample $(N=25)$ of EGs
and lenticular galaxies investigated
within the SAURON project (Cappellari et
al., 2006), and deduced values of the
parameters of interest.   Column caption:
(1) NGC number; (2) effective (half-light)
radius, $R_e$, measured in the $I$-band;
(3) total observed
$I$-band galaxy magnitude; (4)
mass-luminosity ratio (including
DM) deduced from the
best fitting three-integral Schwartzschild
model, computed at a fiducial inclination;
(5) mass-luminosity ratio of the stellar
population; (6) galaxy distance modulus
(hats avoid confusion with the fractional
mass, $m$, and the total mass, $M$); (7)
galaxy mass, calculated as $M_i/{\rm M}_{10}=
(L/{\rm L}_\odot)[(M_i/L)/(10^{10}{\rm m}_\odot/
{\rm L}_\odot)]$; $L/{\rm L}_\odot=\exp_{10}\{-0.4
[I_T-(\hat{m}-\hat{M})-4.11]\}$;  (8) galaxy central
velocity dispersion, deduced from the luminosity-weighted
second moment of the line-of-sight velocity distribution
within the effective radius, $\sigma_e$, as $(\sigma_i)_
{R_e/8}=8^{0.066}\sigma_e$;
(9) galaxy scaling radius,
calculated as $r_i^\dagger/{\rm kpc}=(R_e/{\rm kpc})/
1.81$; $R_e/{\rm kpc}=[(R_e/{\rm arcsec})(d/{\rm Mpc})]/
206.265$; $d/{\rm Mpc}=\exp_{10}[(\hat{m}-\hat{M})
/5-5]$.   The factor, 1.81, is related to an
assumed Hernquist profile for the inner subsystem
(Hernquist, 1990).   The factor, 206.265, is
related to the choice of measure units.    For
further details refer to the parent
paper (Cappellari et al., 2006).}
\label{t:sauron}
\begin{center}
\begin{tabular}{|c|c|c|c|c|c|c|c|c|} \hline
NGC & $R_e$ & $I_T$ & $M/L$ & $M_i/L$ & $(\hat{m}-\hat{M})$
& $M_i$ & $(\sigma_i)_{R_e/8}$ & $r_i^{\dagger}$ \\
& (arcsec) & (mag) & ($I$-band) & ($I$-band) & (mag) & (M$_{10})$
& (km s$^{-1}$) & (kpc) \\
(1)  & (2) & (3)   & (4)  & (5)  & (6)   & (7)   & (8)    & (9)  \\
\hline
0821 & 039.0 & 09.47 & 3.08 & 2.60 & 31.85 & 10.26 & 216.80 & 2.45 \\
2974 & 024.0 & 09.43 & 4.52 & 2.34 & 31.60 & 07.61 & 267.28 & 1.34 \\
3377 & 038.0 & 08.98 & 2.22 & 1.75 & 30.19 & 02.35 & 158.30 & 1.11 \\
3379 & 042.0 & 08.03 & 3.36 & 3.08 & 30.06 & 08.80 & 230.57 & 1.16 \\
3608 & 041.0 & 09.40 & 3.71 & 2.57 & 31.74 & 09.77 & 204.19 & 2.45 \\
4278 & 032.0 & 08.83 & 5.24 & 3.05 & 30.97 & 09.64 & 264.98 & 1.34 \\
4374 & 071.0 & 07.69 & 4.36 & 3.08 & 31.26 & 36.35 & 318.90 & 3.40 \\
4458 & 027.0 & 10.68 & 2.28 & 2.27 & 31.12 & 01.50 & 097.50 & 1.21 \\
4473 & 027.0 & 08.94 & 2.91 & 2.88 & 30.92 & 07.86 & 220.24 & 1.10 \\
4486 & 105.0 & 07.23 & 6.10 & 3.33 & 30.97 & 45.97 & 341.84 & 4.40 \\
4552 & 032.0 & 08.54 & 4.74 & 3.35 & 30.87 & 12.62 & 289.07 & 1.28 \\
4621 & 046.0 & 08.41 & 3.03 & 3.12 & 31.25 & 18.80 & 242.04 & 2.19 \\
4660 & 011.0 & 09.96 & 3.63 & 2.96 & 30.48 & 02.11 & 212.21 & 0.37 \\
5813 & 052.0 & 09.12 & 4.81 & 2.97 & 32.48 & 28.89 & 263.83 & 4.36 \\
5845 & 004.6 & 11.10 & 3.72 & 2.96 & 32.01 & 03.02 & 274.16 & 0.31 \\
5846 & 081.0 & 08.41 & 5.30 & 3.33 & 31.92 & 37.19 & 273.01 & 5.25 \\
\hline
\end{tabular}
\end{center}
\end{table}
More specifically, with regard to
the inner (stellar) subsystem,
projected central
velocity dispersions are
deduced from the luminosity-weighted
second moment of the line-of-sight velocity distribution
within the effective radius, $\sigma_e$, as $(\sigma_i)_
{R_e/8}=8^{0.066}\sigma_e$;
masses are deduced from
luminosities and mass-luminosity
ratios (in $I$-band), as $M_i/
{\rm M}_{10}=(L/{\rm L}_\odot)
[(M_i/L)/(10^{10}{\rm m}_\odot/
{\rm L}_\odot)]$; $L/{\rm L}_
\odot=\exp_{10}\{-0.4[I_T-(\hat
{m}-\hat{M})-4.11]\}$; scaling radii
are calculated from effective radii
(in arcsec) and distances, by use
of a profile factor, equal to 1.81,
related to the case under discussion
(Hernquist, 1990), as $r_i^\dagger/
{\rm kpc}=(R_e/{\rm kpc})/1.81$;
$R_e/{\rm kpc}=[(R_e/{\rm arcsec})(d/{\rm Mpc})]/
206.265$; $d/{\rm Mpc}=\exp_{10}[(\hat{m}-\hat{M})
/5-5]$.   For
further details refer to the parent
paper (Cappellari et al., 2006).

The substitution of the selected value
of the fractional mass, $m$, into
Eq.\,(\ref{eq:sig0i}), allows the
value of the fractional scaling radius,
$y^\dagger$, for each sample object.
Finally, the substitution of $(y^\dagger,
m)$ values into Eq.\,(\ref{eq:phi}),
particularized to HH density profiles
via Eqs.\,(\ref{eq:lumH})-(\ref{seq:welH}),
allows the value of the fractional virial
potential energy, $\phi$, and
the position of each sample object on
a selected isofractional mass curve in
the $({\sf O}y\phi)$ plane, can be
determined.

\begin{figure*}[t]
\begin{center}
\includegraphics[scale=0.8]{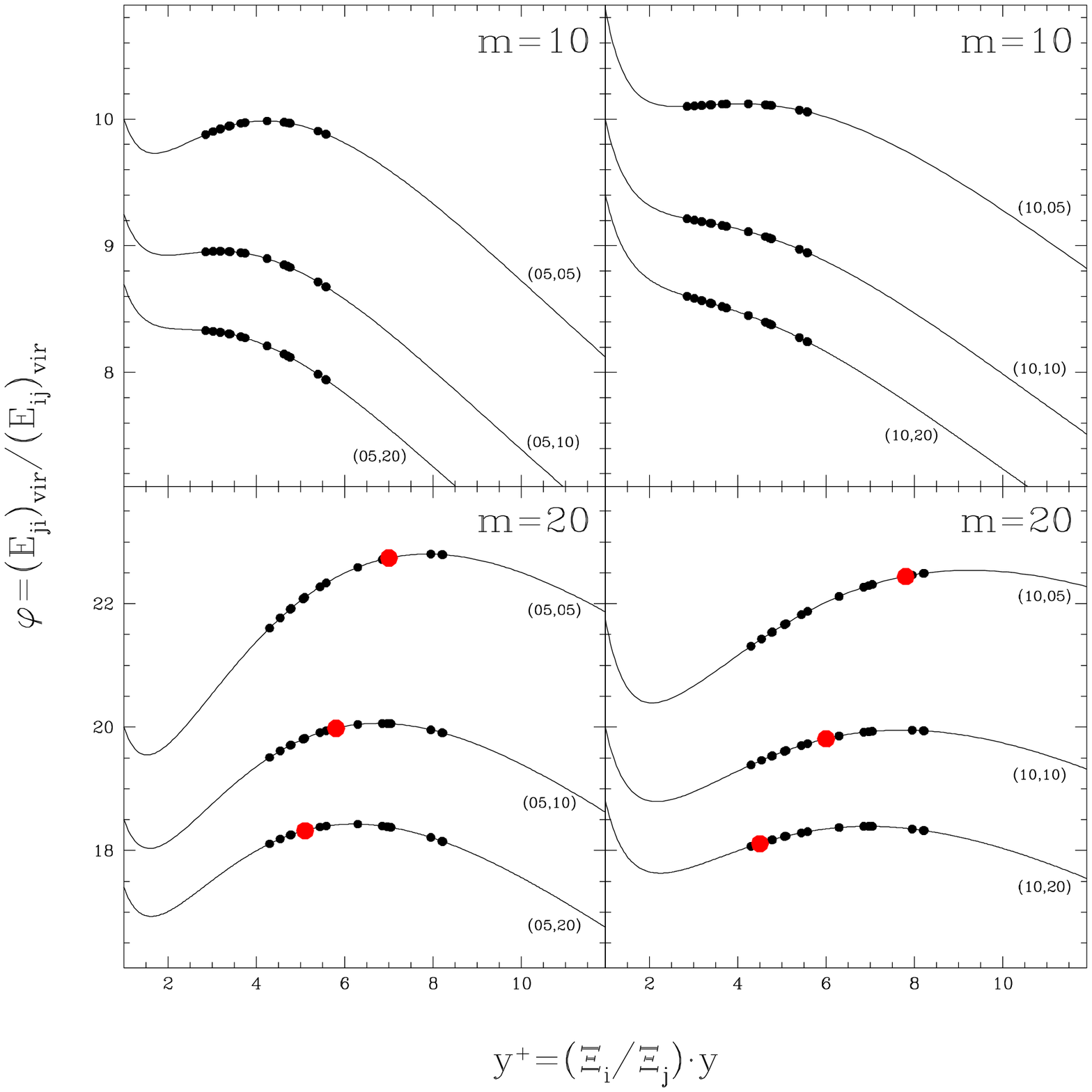}
\caption{Isofractional mass curves, $m=10$
(top panels), and $m=20$ (bottom panels),
related to HH macrogases, for different choices
of scaled truncation radii, $(\Xi_i,
\Xi_j)$, indicated near the corresponding curves,
and related positions of $N=16$ EGs
(dots), listed in Tab.\,\ref{t:sauron}.
Different cases correspond to vertical shifts of
the sample objects.   Larger dots
%(red in the online version of the paper)
represent configurations
where the virial potential energy of the inner subsystem
attains the maximum value with respect to a frozen
outer subsystem.}
\label{f:HHm}
\end{center}
\end{figure*}

The results are plotted for different
choices of scaled truncation radii,
$(\Xi_i,\Xi_j)$, and different choices
of fractional masses, $m$, in
Fig.\,\ref{f:HHm}.
Isofractional mass curves, $m=10$
(top panels), and $m=20$ (bottom panels),
are labelled by the selected choices
of scaled truncation radii, $(\Xi_i,
\Xi_j)$, indicated near the corresponding curves,
and the sample objects are represented as dots.
Curves lie above and below the critical curve
for $m=10$, while all curves lie above the
critical curve for $m=20$, with respect to
the cases considered.

The above description is due to reasons
of simplicity, but the model does not
necessarily imply that sample objects
must be located on the same  isofractional
mass curve and/or correspond to the same
scaled truncation radii.   What is
relevant is the position of EGs
on the $({\sf O}y\phi)$ plane, which
implies the following assumptions: (i)
there are in the universe 16 EGs with
intrinsic values of the parameters
equal to their counterparts listed in
Tab.\,\ref{t:sauron} for sample
objects, and (ii) related star distributions
are well described by H density profiles.

In fact, observational uncertainties
on the quantities of interest (Cappellari
et al., 2006) make large errors on the
fractional scaling radius, $y^\dagger$.
As an example, a single galaxy, NGC 3379,
and a single parameter, the effective
radius, shall be considered.    In the
case under discussion, $R_e=(42.0\mp7.1)$
arcsec (Cappellari et al., 2006), but
different estimates exist, such as
$(54.8\mp3.5)$ arcsec (Capaccioli et al.,
1990) which is consistent with the above
result within $2\sigma_{R_e}$.   Using
the latter value, the fractional scaled
radius, $y^\dagger$, passes from  5.74
to 4.08 for a fractional mass, $m=10$,
and from 8.45 to 6.07 for $m=20$.

\begin{figure*}[t]
\begin{center}
\includegraphics[scale=0.8]{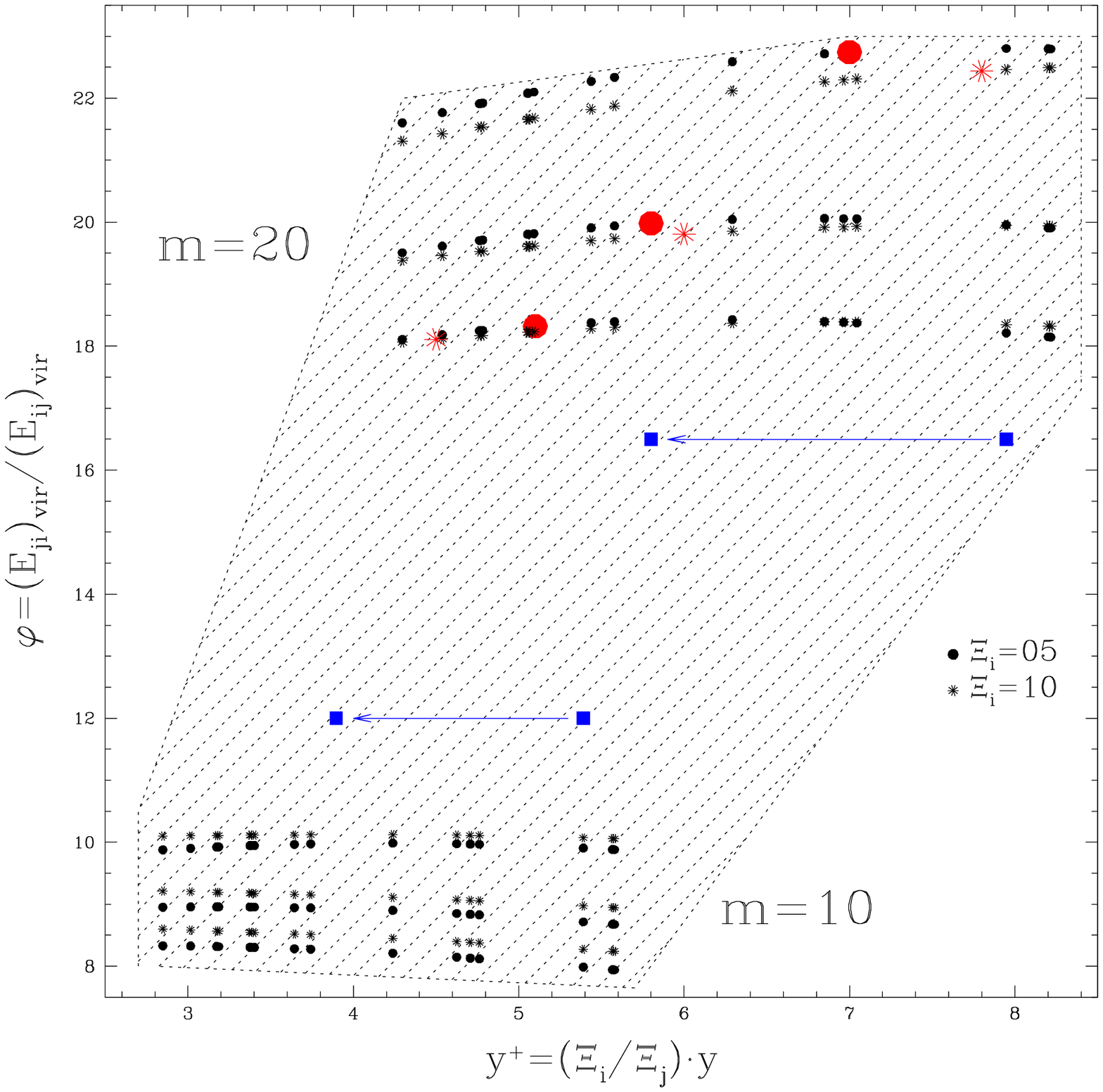}
\caption{The parameter space of the cases plotted
in Fig.\,\ref{f:HHm}, roughly outlined by the shaded
region.   Dots and asterisks correspond to inner scaled
truncation radii, $\Xi_i=5$ and 10, respectively.   For
fixed fractional mass, $m$, the corresponding outer
scaled truncation
radii read $\Xi_j=5, 10, 20$, from top to bottom.   Small
and large symbols represent sample objects and configurations
where the virial potential energy of the inner subsystem
attains the maximum value with respect to a frozen
outer subsystem, respectively.   No such
configuration exists in the cases
considered, for fractional masses, $m\appleq20$.
The related parameter space is restricted
to a region close to the angle defined
by the locus of large dots and asterisks,
respectively.    The intersection between
the above mentioned loci occurs for outer
scaled truncation radii, $\Xi_j$, slightly
larger than 10.   The change of position
for NCG 3379, due to a change in effective
radius, from $R_e=42.0$ to $R_e=54.8$, is
indicated by squares, regardless of
the vertical scale.}
\label{f:HHs}
\end{center}
\end{figure*}

Different
values of parameters, $m$, $\Xi_i$, $\Xi_j$,
makes sample objects locate on different
isofractional mass curves, as shown in
Fig.\,\ref{f:HHm}, where $y$ depends on
$m$ but is independent of $\Xi_i$ and
$\Xi_j$.   Fiducial values of the above
mentioned parameters, say $10\le m\le20$,
$5\le\Xi_u\le20$, $u=i,j$, define a
region on the $({\sf O}y\phi)$ plane,
approximately as: $2.5<y^\dagger<6.0$,
$7<\phi<11$, $m=10$; $4<y^\dagger<9$,
$17<\phi<23$, $m=20$; where sample
objects are positioned, as shown in
Fig.\,\ref{f:HHs}.
Dots and asterisks correspond to inner scaled
truncation radii, $\Xi_i=5$ and 10, respectively.   Small
and large symbols represent sample objects and configurations
where the virial potential energy of the inner subsystem
attains the maximum value with respect to a frozen
outer subsystem, respectively.   No such
configuration exists in the cases
considered, for fractional masses, $m\appleq20$.
The related parameter space is restricted
to a region close to the angle defined
by the locus of large dots and asterisks,
respectively.    The intersection between
the above mentioned loci occurs for outer
scaled truncation radii, $\Xi_j$, slightly
larger than 10.
The change of position
for NCG 3379, due to a change in effective
radius, from $R_e=42.0$ to $R_e=54.8$, is
indicated by squares, regardless of
the vertical scale.

In the special situation where the outer
subsystem remains frozen and the inner
one is free to contract or to expand,
with unchanged mass and density profile,
the virial potential energy of the inner subsystem
may attain an extremum point of maximum
(Secco, 2000, 2001, 2005; Marmo and Secco,
2003).   For HH density profiles, lower
fractional masses $(m=10)$ yield no
extremum point, while the contrary holds
for larger fractional masses $(m=20)$.
The special configurations related to
the maximum virial potential energy of the inner
subsystem, when the outer subsystem
remains frozen (hereafter quoted as
``the maximum configuration''), are
represented as large dots
%(red in the online version of the paper)
in Fig.\,\ref{f:HHm} (bottom panels)
and large dots and asterisks in
Fig.\,\ref{f:HHs}.

The maximum configuration appears to
have little relevance in the light of
the current model for a number of
reasons.   First, it occurs for
fractional masses above a threshold.
Second, the loci of maximum configurations
on the $({\sf O}y\phi)$ plane show
no correlation with the parameter
space of the model related to sample
objects.   More
precisely, the loci of maximum
configurations are narrower and
extend from bottom left to top
right, while sample objects lie
on a broader and less inclined
band, as shown in Fig.\,\ref{f:HHs}.

The position of the maximum
configuration
on the isofractional mass curves,
$m=20$, depends on the concentration
of the outer subsystem, while the
position of sample objects can
only be vertically shifted, as the
fractional scaling radius, $y^\dagger$,
is independent of the concentration,
via Eq.\,(\ref{eq:sig0i}).
In this view, it is difficult to conceive
maximum configurations as ending
points of any evolutionary
track on the $({\sf O}y\phi)$ plane,
even if observational uncertainties
are high.

EG position on the $({\sf O}y\phi)$ plane,
in particular along an isofractional mass
curve $(m={\rm const})$, represents the
ending point of evolutionary
tracks on the above mentioned plane.
The related configurations may be
thought of as virialized to a good extent,
in that sample objects listed in Tab.\,\ref
{t:sauron} show no sign of ungoing merger
and star formation burst.   The coincidence of
ending points with maximum configurations
would imply, for fixed $(m, \Xi_i,\Xi_j)$:
(i) a single ending point for all sample
objects i.e. homologous evolution, and
(ii) $m$ above the threshold
for the occurrence of maximum configurations,
which does not necessarily happen, as shown
in Fig.\,\ref{f:HHm}.   On the contrary, the
evolution of EGs
appears to be non homologous, in the
sense that different ending points
take place 
along the selected isofractional mass
curves defined by the HH macrogases equation
of state.

Owing to Eqs.\,(\ref{eq:Eitb}) and
(\ref{eq:vir2b}), the total
energy of the virialized system reads:
\begin{equation}
\label{eq:EE}
E=\frac12[(E_{ij})_{\rm vir}+(E_{ji})_{\rm vir}]=-(E_i)_{\rm kin}-
(E_j)_{\rm kin}~~;
\end{equation}
and the combination of Eqs.\,(\ref{eq:vir2b}),
(\ref{eq:phic}), and (\ref{eq:EE}) yields:
\begin{equation}
\label{eq:Ephi}
E=\frac12(1+\phi)(E_{ij})_{\rm vir}=-(1+\phi)(E_i)_{\rm kin}~~;\qquad
\phi=\frac{(E_j)_{\rm kin}}{(E_i)_{\rm kin}}~~;
\end{equation}
where the kinetic energy of the inner
subsystem, $(E_i)_{\rm kin}$, may be
deduced from observations, and the
related total energy, $E$, may be
read on the vertical axis of Fig.\,\ref
{f:HHm}.   Within the range, $\phi_{\rm
min}\le\phi\le\phi_{\rm max}$, three
configurations exist with same virial
potential (or kinetic) energy ratio, $\phi$,
and fractional mass, $m$, but different
fractional truncation radius, $y$.

The restriction to constant fractional mass
also in time, makes evolutionary tracks on
the $({\sf O}y\phi)$ plane locate on the
related isofractional mass curve, 
$(m={\rm const})$, where displacements
from the left to the right (increasing
$y$ values) are due to energy dissipation,
and displacements
from the right to the left (decreasing
$y$ values) are due to energy acquisition.
In the former alternative, 
changes in $\phi$ and $(E_{ij})_{\rm
vir}$ must act to yield a decreasing
$E$ (increasing in absolute value),
the larger energy change being related
to the larger $y$ change.   For further
details refer to Appendix \ref{a:qusco}.
In this view,
the maximum configuration corresponds
to a special energy change, $(\Delta E)_
{\rm max}$, and to a special $y$ change,
$(\Delta y)_{\rm max}$.   But there is no
apparent reason for which a special amount
of energy has to be dissipated starting
from the beginning of evolution,
to attain the maximum configuration.
A similar result holds in the latter
alternative.   In fact, it can be
seen in Fig.\,\ref{f:HHm} that the
sample objects show no connection
with the maximum configuration.
Then the maximum configuration,
which implies a frozen outer subsystem,
has no special relevance
in the light of the current model.
%On the other hand, it is worth
%remembering that earlier investigations
%on the maximum configuration
%(Secco, 2000, 2001, 2005; Marmo and
%Secco, 2003) were performed in dealing
%with a transition from a turnaround
%configuration where gas exhaustion via
%star formation had previously occurred,
%to a virialized state with no energy
%dissipation.
For an analysis
of different theories on the maximum
configuration, refer to a specific study
(Valentinuzzi 2006, Chaps.\,3-5).

The above results hold if the baryonic
subsystem is mainly in form of stars.   Let
us take into consideration a different
scenario, where a less concentrated gaseous subsystem
than the stellar one is also present,
as in the Coma cluster of galaxies
(e.g., Briel et al., 1992), and assume the same
mass distribution as in the non baryonic
matter to preserve use of two-component
models (Caimmi, 2003).   Accordingly, a
typical EG is idealized
as formed by an inner subsystem
made of stars and an outer subsystem made
of gas and non baryonic matter.   As the
amount of baryonic and non baryonic
matter have to remain unchanged, the
inner and the outer subsystem are less
and more massive, respectively, than
in absence of undetected baryons.   Again,
it is assumed that the related mass
distributions are represented by HH
density profiles.     Then
a similar procedure may be followed,
keeping in mind that the fractional
mass, $m$, is the ratio of nonbaryonic
+ extragalactic gas mass to star mass.
For further details refer to an earlier
attempt (CM03).

\section{Conclusion}\label{conc}

Two-component systems
have been conceived as
(two-component) macrogases, and the
related equation of state has been
formulated using the virial theorem
for subsystems (Limber, 1959; Brosche
et al., 1983; Caimmi et al., 1984;
Caimmi and Secco, 1992), under the
restriction of (i) homeoidally
striated ellipsoids (Roberts, 1962)
and (ii) similar and similarly
placed boundaries.

Explicit calculations have been
performed for a useful reference
case and a few cases of astrophysical
interest, both in presence and in
absence of truncation radius.
More specifically, the following
cases have been dealt with: IJ=
UU, PP, HH, HP, HN, where I and J
denote the inner and the outer
density profile, respectively,
and the other captions relate
to the following density profiles:
U ($\rho=$const), P (Plummer, 1911),
H (Hernquist, 1990), N (Navarro et
al., 1995, 1996, 1997).   Shallower
density profiles (UU, PP), have
been found to yield an equation of
state, $\phi=\phi(y,m)$, characterized
by the occurrence of two extremum
points, one maximum and one minimum,
as in an earlier attempt (CS90).
Steeper density profiles (HH, HP,
HN), have been found to produce a
similar equation of state where,
in addition, a single horizontal
inflexion point occurs in a critical
isofractional mass curve, and
isofractional mass curves related
to lower values, $m=M_j/M_i<m_{\rm
crit}$, show no extremum point.
The similarity between isofractional
mass curves and van der Waals'
isothermal curves, has suggested
the possibility that a phase
transition could take place in a
bell-shaped region of the $({\sf O}y\phi)$ plane.

Further investigation has been devoted to HH
density profiles for which a numerical
algorithm (Ciotti et al., 1996) has been used to
represent EGs
and their hosting DM haloes along selected
isofractional mass curves on the $({\sf O}y\phi)$
plane, under the assumption that the related
fractional mass has the same value in different
systems.    In the light of the model, the
evolution of EGs
has been found to be non strictly homologous,
in the sense that the end of evolutionary
tracks on the $({\sf O}y\phi)$ plane
occur at different points along
the related isofractional mass curve, instead
of being close to a single point.

\section{Acknowledgements}
We are indebted to an anonymous referee for critical
comments which  improved an
earlier version of the manuscript.
Thanks are due to L. Secco for fruitful discussions.
The analytical integrations needed in the current paper
were helped substantially by use of the Mathematica
package and visiting the internet site:
``HTTP://INTEGRALS. WOLFRAM.COM/INDEX.CGI''.   This
is why we are deeply grateful to the Wolfram staff,
in particular to Daniel Lichtblau, and wish to
acknowledge all the facilities encountered
therein.

\appendix
\section*{Appendix}

\section{Plummer density profiles and $n=5$
polytropes}\label{a:pol5}

The Lane-Emden equation reads (e.g.,
Lane, 1870; Emden, 1907; Chandrasekhar,
1933; Horedt, 2004, Chap.\,2, \S2.1):
\begin{leftsubeqnarray}
\slabel{eq:LEa}
&& \frac1{\xi_{\rm LE}^2}\frac\diff{\diff\xi_{\rm LE}}\left(\xi_{\rm LE}^2
\frac{\diff\theta}{\diff\xi_{\rm LE}}\right)=-\theta^n~~; \\
\slabel{eq:LEb}
&& \theta(0)=1~~;\qquad\theta(\Xi_{\rm LE})=0~~; \\
\slabel{eq:LEc}
&& \xi_{\rm LE}=\frac r{\alpha_{\rm LE}}~~;\qquad\Xi_{\rm LE}=\frac R
{\alpha_{\rm LE}}~~; \\
\slabel{eq:LEd}
&& \rho(r)=\lambda\theta^n(\xi_{\rm LE})~~;
\label{seq:LE}
\end{leftsubeqnarray}
where $n$ is the polytropic index
($0\le n\le5$ and $0.5<n\le5$ for
realistic collisional and collisionless
equilibrium configurations, respectively),
$\alpha_{\rm LE}$ is a scaling radius and
$\lambda$ is the central density.

In the special case, $n=5$, the integration
of Eq.\,(\ref{eq:LEa}) yields (Schuster,
1883; see also Chandrasekhar, 1939,
Chap.\,IV, \S4; Horedt, 2004, Chap.\,2,
\S2.3.3):
\begin{leftsubeqnarray}
\slabel{eq:LE5a}
&& \theta(\xi_{\rm LE})=\left(1+\frac13\xi_{\rm LE}^2\right)^{-1/2}~~;
\qquad\Xi_{\rm LE}\to+\infty~~;\qquad n=5~~; \\
\slabel{eq:LE5b}
&& \frac{\diff\theta}{\diff\xi_{\rm LE}}=-\frac13\xi_{\rm LE}\left(1+
\frac13\xi_{\rm LE}^2\right)^{-3/2}~~; \\
\slabel{eq:LE5c}
&& -\xi_{\rm LE}^2\frac{\diff\theta}{\diff\xi_{\rm LE}}=\sqrt{3}\left(\frac
{\xi_{\rm LE}}{\sqrt{3}}\right)^3\left[1+\left(\frac{\xi_{\rm LE}}
{\sqrt{3}}\right)^2\right]^{-3/2}~~; \\
\slabel{eq:LE5d}
&& \lim_{\xi_{\rm LE}\to+\infty}\left(-\xi_{\rm LE}^2\frac{\diff\theta}
{\diff\xi_{\rm LE}}\right)=\sqrt{3}~~;
\label{seq:LE5}
\end{leftsubeqnarray}
and the general expression of a polytrope mass
(e.g., Chandrasekhar, 1939, Chap.\,IV, \S4;
Horedt, 2004, Chap.\,2, \S2.6.3):
\begin{equation}
\label{eq:LEM}
M=-4\pi\lambda\alpha_{\rm LE}^3\Xi_{\rm LE}^2\left(\frac{\diff\theta}
{\diff\xi_{\rm LE}}\right)_{\Xi_{\rm LE}}~~;
\end{equation}
in the case of interest, using
Eqs.\,(\ref{seq:LE5}), reduces to:
\begin{equation}
\label{eq:LEM5}
M=\sqrt{3}\,4\pi\lambda\alpha_{\rm LE}^3~~;\qquad n=5~~;
\end{equation}
finally, the explicit expression of the
density profile results from the
combination of Eqs.\,(\ref{eq:LEd})
and (\ref{eq:LE5a}), as:
\begin{equation}
\label{eq:LErh5}
\rho(r)=\lambda\left(1+\frac13\xi_{\rm LE}^2\right)^{-5/2}~~;\qquad n=5~~;
\end{equation}
and the following relation is derived from
comparison with Eqs.\,(\ref{eq:rhoa}) and
(\ref{eq:ex3}):
\begin{equation}
\label{eq:LE5P}
\rho^\dagger2^\chi\xi^{-\gamma}(1+\xi^\alpha)^{-\chi}=\lambda\left(1+
\frac13\xi_{\rm LE}^2\right)^{-5/2}~~;
\end{equation}
which, in turn, implies the following:
\begin{leftsubeqnarray}
\slabel{eq:LE5Za}
&& \xi=\frac{\xi_{\rm LE}}{\sqrt{3}}~~;\qquad\alpha_{\rm LE}=\frac\xi
{\xi_{\rm LE}}r^\dagger=\frac{r^\dagger}{\sqrt{3}}~~; \\
\slabel{eq:LE5Zb}
&& \gamma=0~~;\qquad\alpha=2~~;\qquad\chi=\frac52=\frac{\beta-\gamma}\alpha=
\frac\beta2~~;\qquad\beta=5~~; \\
\slabel{eq:LE5Zc}
&& \lambda=2^{5/2}\rho^\dagger~~;
\label{seq:LE5Z}
\end{leftsubeqnarray}
and the substitution of Eqs.\,(\ref{eq:LE5Za})
and (\ref{eq:LE5Zc}) into (\ref{eq:LErh5})
yields:
\begin{leftsubeqnarray}
\slabel{eq:ZPa}
&& \rho(r)=\rho^\dagger f(\xi)~~; \\
\slabel{eq:ZPb}
&& f(\xi)=\frac{2^{5/2}}{(1+\xi^2)^{5/2}}~~;
\label{seq:ZP}
\end{leftsubeqnarray}
according to Eq.\,(\ref{eq:fP}).

An equivalent formulation can be obtained
by the combination of Eqs.\,(\ref{eq:LEM5})
and (\ref{seq:LE5Z}).   The result is:
\begin{equation}
\label{eq:P}
\rho(r)=\frac{3M}{4\pi}\frac{(r^\dagger)^2}{[r^2+(r^\dagger)^2]^{5/2}}~~;
\end{equation}
which is known as the Plummer (1911)
density profile.

\section{Quasi static contraction in presence of
tidal potential}\label{a:qusco}

The scalar virial equations, Eqs.\,(\ref
{eq:vir2b}), for assigned density profiles
i.e. fixed $\alpha_u$, $\beta_u$, $\gamma_u$,
and $\Xi_u$, $u=i,j$, depend on four parameters
e.g., masses and scaling radii, $M_u$ and
$r_u^\dagger$.   The further assumption of
spherical-symmetric matter distributions is
only to simplify calculations.   Let an
assigned amount of energy, $\Delta E<0$,
be instantaneously dissipated within the inner
subsystem, as:
\begin{equation}
\label{eq:DE}
(E_i)_{\rm kin}\to(E_i)_{\rm kin}+\Delta E~~;
\end{equation}
and the system readjust to attain
a new configuration in accordance with Eqs.\,(\ref
{eq:vir2b}) where, in general, density profiles
and scaling radii are changed, while masses and
the outer subsystem truncation radius may be
assumed fixed, or their variations assigned.
More specifically,
density profiles are steepened by energy dissipation,
and the system gets bounder i.e. a larger (in
absolute value) binding energy is attained.

The further constraint of density profiles
unaffected by energy dissipation, implies the
following changes $(u=i,j)$:
\begin{leftsubeqnarray}
\slabel{eq:Dpa}
&& M_u\to M_u+\Delta M_u~~;\qquad\Delta M_u=0~~; \\
\slabel{eq:Dpb}
&& \Xi_u\to\Xi_u+\Delta\Xi_u~~;\qquad\Delta\Xi_u=0~~; \\
\slabel{eq:Dpc}
&& R_u\to R_u+\Delta R_u~~;\qquad\Delta R_j=0~~; \\
\slabel{eq:Dpd}
&& r_u^\dagger\to r_u^\dagger+\Delta r_u^\dagger~~;\qquad\Delta r_j^
\dagger=0~~; \\
\slabel{eq:Dpe}
&& (E_u)_{\rm kin}\to(E_u)_{\rm kin}+\Delta E_u~~;\qquad\Delta E_i+
\Delta E_j=\Delta E~~;
\label{seq:Dp}
\end{leftsubeqnarray}
where the assumption that energy dissipation
within the inner subsystem has no effect on
(i) masses, $M_u$; (ii) scaled truncation radii,
$\Xi_u$; density profiles, $(\alpha_u,\beta_u,
\gamma_u)$; (iv) truncation radius of the
outer subsystem, $R_j$; implies a frozen outer
subsystem.
In general, the changes, $\Delta M_u=\zeta_m
M_u$, $\Delta R_u=\zeta_uR_u$, and $\Delta
r_u^\dagger=\zeta_ur_u^\dagger$, which make
$m$ and $\Xi_u$ conserved, should
be specified.
The application of the scalar virial theorem,
Eqs.\,(\ref{seq:vir2}), to the subsystems
before and after energy dissipation, reads:
\begin{leftsubeqnarray}
\slabel{eq:Ua}
&& U_{uv}(y^\dagger)+2(E_u)_{\rm kin}=0~~; \\
\slabel{eq:Ub}
&& U_{uv}(y^\dagger+\Delta y^\dagger)+2[(E_u)_{\rm kin}+\Delta E_u]=0~~; \\
\slabel{eq:Uc}
&& U_{uv}=(E_{uv})_{\rm vir}~~;\qquad u=i,j~~;\qquad v=j,i~~;
\label{seq:U}
\end{leftsubeqnarray}
where the fractional scaling radius, $y^
\dagger=r_j^\dagger/r_i^\dagger$, has been
chosen as variable.

The combination of Eqs.\,(\ref{eq:Ua}) and
(\ref{eq:Ub}) yields:
\begin{leftsubeqnarray}
\slabel{eq:DEva}
&& \Delta(E_{uv})_{\rm vir}+2\Delta E_u=0~~;\qquad u=i,j~~;\qquad v=j,i~~; \\
\slabel{eq:DEvb}
&& \Delta(E_{uv})_{\rm vir}=U_{uv}(y^\dagger+\Delta y^\dagger)-
U_{uv}(y^\dagger)~~;
\label{seq:DEv}
\end{leftsubeqnarray}
and the combination of Eqs.\,(\ref{eq:DEva}), by
use of (\ref{eq:Dpe}) produces:
\begin{equation}
\label{eq:D3E}
\Delta(E_{ij})_{\rm vir}+\Delta(E_{ji})_{\rm vir}+2\Delta E=0~~;
\end{equation}
which is a transcendental equation in
$\Delta y^\dagger$ provided the density
profiles and the amount of dissipated
energy, $\Delta E$, are specified.
In general, the changes, $\Delta M_u=\zeta_m
M_u$, $\Delta R_u=\zeta_uR_u$, and $\Delta
r_u^\dagger=\zeta_ur_u^\dagger$, which make
$m$ and $\Xi_u$ conserved, should also
be specified.
Then the remaining parameters related
to the relaxed system, due to energy
dissipation, may be determined.   In
particular, $\Delta y^\dagger>0$
is expected together with $m={\rm const}$
due to mass conservation in each subsystem
or mass variation of the kind considered,
$m=(M_j+\Delta M_j)/(M_i+\Delta M_i)=
M_j/M_i$.
Accordingly, the position of the system
in the $({\sf O}y\phi)$ plane moves from
the left to the right along the related
isofractional mass curve (see e.g.,
Fig.\,\ref{f:HHm}).

%The above procedure can also be applied
%in a different context, where matter is
%exchanged between subsystems, with both
%mass and energy conservation,
%unaffected density profiles,
%$(\alpha_u$, $\beta_u$, $\gamma_u)$,
%and scaled truncation radii,
%$\Xi_u$, $u=i,j$.

\end{document}